\documentstyle[11pt]{article}
\begin{flushright}
{NORDITA--93/29 N}
{SUNY--NTG--92--33}
\end{flushright}
\bigskip
\renewcommand{\theequation}{\thesection.\arabic{equation}}
\newcommand\be{\begin{eqnarray}}
\newcommand\ee{\end{eqnarray}}
\def\ni{\noindent}
\def\Tr{{\,\rm Tr}}
\def\del{\partial}
\def\L{{\cal L}}
\def\su2xsu2{{SU(2)\times SU(2)}}
\def\su3xsu3{{SU(3)\times SU(3)}}

\newcommand {\sol} { $1.44 M_\odot$ }
\begin{document}
\setlength{\baselineskip}{15.2pt}
\parindent=20pt
\begin{center}
{\Large {Composition, structure and evolution of neutron stars \\ with
kaon condensates} }
\vskip 12pt
{\large Vesteinn Thorsson$^a$, Madappa Prakash$^b$ and James. M. Lattimer$^c$ }
\vskip 12pt
$^a$NORDITA \\ Blegdamsvej 17, DK--2100 Copenhagen \O, Denmark
\vskip 12pt
$^b$Physics Department \\ State University of New York at Stony Brook \\
 Stony Brook, NY 11794--3800, USA \\
\vskip 12pt
$^c$Earth and Space Sciences Department\\ State University of New York at
Stony Brook \\
Stony Brook, NY 11794--2100, USA \\
\vskip 12pt
{\bf Abstract} \\
\end{center}
{\small
\begin{quotation}
We investigate the possibility of kaon condensation  in the dense interior of
neutron stars through the s--wave interaction of kaons with nucleons.   We
include nucleon--nucleon interactions by using simple parametrizations of
realistic forces,  and include electrons and muons in $\beta$--equilibrium.
The equation of state above the condensate threshold is derived in the mean
field approximation.   The conditions under which kaon condensed cores undergo
a transition to quark matter containing strange quarks are also established.

The critical density for kaon condensation lies in the range
(2.3--5.0)$~\rho_0$, where $\rho_0=0.16$~fm$^{-3}$ is the
equilibrium density of nuclear matter.  The critical density depends largely on
the value of the strangeness content of the proton, the size of which
is controversial.  For too large a value of the strangeness
content, matter with a kaon condensate is not sufficiently stiff to
support the lower limit of \sol for a neutron star.  Kaon
condensation dramatically increases the proton abundance of matter and
even allows positrons to exist inside the core.

We also consider the case when neutrinos are trapped in the matter, a
situation that applies to newly-formed neutron star matter that is
less than about 10 seconds old.  Neutrino trapping shifts both kaon
condensation and the quark matter transition to higher densities than
in the case of cold, catalyzed matter.  A newly--formed neutron star is
expected to have a rather low central density, the density rising only
after mass accretion onto the star ends after a few seconds.  Thus, it
is likely that if kaon condensation and/or the quark--hadron phase
transition occur, they do so only during or after the mass accretion
and neutrino trapping stages.  We suggest that neutrino observations
from a galactic supernova may provide direct evidence for or against a
condensate and/or a quark--hadron transition.
\end{quotation}}
\vfill
\eject
\bigskip
\bigskip
\setcounter{section}{1}
\setcounter{equation}{0}
\centerline {{\bf 1. Introduction}}
\bigskip
\bigskip

The idea that the ground state of baryonic matter might contain a
Bose--Einstein condensate of kaons above a certain matter density is due to
Kaplan and Nelson~\cite{kapnel}.   Using an $SU(3) \times SU(3)$ chiral
lagrangian, these authors showed that at $\rho \simeq 3 \rho_0 $ where $\rho_0
= 0.16 ~{\rm fm}^{-3}$ is the equilibrium nuclear matter density, matter
containing a kaon condensate was energetically favorable, and, that the
formation of such a state was to a large extent driven by the kaon--nucleon
sigma term $\Sigma^{KN}$. The ansatz for the condensate included pion and kaon
condensates with colinear momenta of arbitrary magnitude. To determine the
ground state, the authors carried out numerically a multivariate minimization
with respect to the momenta, the condensate amplitudes and the charge chemical
potential (to implement charge neutrality).

Brown, Kubodera and Rho~\cite{bkr} came to similar conclusions, using a reduced
$SU(2) \times SU(2) $ model, the V-spin $\sigma$--model, with a symmetry
breaking term given by $\Sigma^{KN}$. In this model, it is assumed that pions
will not condense at the relevant densities, and that interactions other than
s--wave kaon--baryon interactions are relatively small and may be ignored. The
authors also suggested that the condensation of kaons may be understood as a
chiral rotation away from a V--spin scalar ground state. The term in the
Hamiltonian which breaks chiral symmetry explicitly is proportional to
$\Sigma^{K N}( \cos \theta - 1) $, where  $\theta$ is the rotation angle of the
expectation value of the kaon field away from the scalar ground state. After
kaon condensation sets in at a given matter density, the angle $\theta$
increases and the explicit chiral symmetry breaking term is reduced. This will
lead to a partial restoration of the chiral symmetry explicitly broken in the
vacuum.

Politzer and Wise~\cite{polwise} have recently reanalyzed the kaon condensation
threshold in the Kaplan--Nelson model, including both pions  and kaons
interacting with baryons through  s--wave and p--wave interactions. Analytic
expressions for the threshold densities for pion and kaon condensation were
obtained by investigating the energy density for small values of the condensate
amplitude.

Recently Brown, Kubodera, Rho and Thorsson~\cite{bkrt} examined kaon
condensation due to s--wave meson--baryon interactions in the  Kaplan--Nelson
model.  The analyses included nucleon--nucleon interactions  from  a simple
parametrization due to Prakash, Ainsworth and Lattimer~\cite{pal} and included
electrons in $\beta$-equilibrium with the hadronic  matter.
In addition to the dependence on $\Sigma^{K N}$, the condensation
threshold also depends on the nuclear symmetry energy.    The energy density
and composition of the ground state above the condensation threshold was also
studied.  A basic premise of this work was that above the condensation
threshold, it becomes energetically favorable for kaons to be converted  into
electrons through strangeness changing reactions. A simple
estimate indicated that these reactions take place rapidly enough to establish
chemical equilibrium during neutron star  cooling. It was found that the
symmetry energy favors relatively large values of the proton to  baryon ratio
$x$.   The large proton fraction $x$ found implies that kaon condensation
might trigger the rapid cooling of neutron stars via the direct URCA process
involving nucleons~\cite{lpph}.  In the treatment of ref.~\cite{bkrt}, the
formation of a kaon condensate does not require the existence of a pion
condensate as in ref.~\cite{polwise}.  It was considered likely that the in
medium modification of  $g_A$, the axial vector coupling  in the nuclear
medium, effectively prevents the condensation of pions.

In this work, we improve and extend the analysis of ref.~\cite{bkrt}
and discuss its astrophysical implications.  The work is organized as
follows.  In sect. 2, we establish the equilibrium conditions of
matter in beta equilibrium in the presence of a kaon condensate.  The
threshold density for kaon condensation and the composition of matter
above the condensation threshold are also discussed here.  In sect. 3,
we construct neutron stars with a kaon condensate.  The structural
changes brought about by the condensate and the implications thereof
are then presented.  Whether the inner cores of such stars can contain
quark matter or not is investigated in sect. 4.  Section 5 contains a
discussion of the situation when neutrinos are trapped.  We also
discuss the evolution of a newly-formed neutron star from the
neutrino-trapped state to the final cold, catalyzed state.  Avenues
for future investigation are outlined in section on 6. A brief
discussion of the kaon self--energy and the condensation threshold is
given in appendix A.

\bigskip
\bigskip
\addtocounter{section}{1}
\setcounter{equation}{0}
{\centerline {\bf 2. S--wave kaon condensation in the Kaplan--Nelson model}}
\bigskip
\bigskip

{\ni {2.1. INTERACTIONS }}
\bigskip
\bigskip

Our starting point is the effective chiral Lagrange density introduced by
Kaplan and Nelson~\cite{kapnel} that involves an octet of pseudoscalar mesons
(pseudo-Goldstone bosons) $M$ and an octet of baryons $B$
\be
\L &=& \frac{f^2}{4}\Tr \del_\mu U\del^\mu U^\dagger + c\Tr \left( m_q(U
+U^\dagger - 2) \right) \nonumber\\
&+& \Tr \bar{B} \gamma^\mu i\del_\mu B+i\Tr B^\dagger [V_0, B]
-D\Tr B^\dagger\vec{\sigma}\cdot \{\vec{A}, B\}-F\Tr B^\dagger \vec{\sigma}
\cdot [\vec{A}, B]\nonumber\\
&+&a_1\Tr B^\dagger\left(\xi m_q \xi+ h.c.\right) B +a_2\Tr B^\dagger B
\left(\xi m_q\xi
+ h.c.\right) +a_3\Tr B^\dagger B \Tr\left(m_qU + h.c.\right)\nonumber\\
&+& \L_e  \,\,\,+\,\,\, \L_\mu\,,
\label{kapnel}
\ee
with the usual non--linear sigma field $U$ and the $\xi$ field given
by
\be
U = \exp{(\sqrt{2}iM/f)} \,\,\, , \,\,\xi^2=U
\ee
and, the mesonic vector and axial vector currents by
\be
V_\mu = \frac12 ( \xi^\dagger \del_\mu \xi + \xi \del_\mu \xi^\dagger )\,,\,
A_\mu = \frac{i}{2} ( \xi^\dagger \del_\mu \xi - \xi \del_\mu \xi^\dagger )\,.
\ee
The meson and baryon octets are, respectively,
\be
M = \left[ \begin{array}{ccc}
\frac{1}{\sqrt{2}}\pi^0+\frac{1}{\sqrt{6}}\eta &  \pi^+ & K^+  \\
\pi^-  & -\frac{1}{\sqrt{2}}\pi^0+\frac{1}{\sqrt{6}}\eta  &  K^0  \\
 K^-   &  \bar{K}^0 & - \sqrt{\frac23}\eta
    \end{array}  \right]
\label{matrixm}
\ee
and
\be
B = \left[ \begin{array}{ccc}
\frac{1}{\sqrt{2}}\Sigma^0+\frac{1}{\sqrt{6}}\Lambda &  \Sigma^+ & p  \\
\Sigma^-  & -\frac{1}{\sqrt{2}}\Sigma^0+\frac{1}{\sqrt{6}}\Lambda  &  n  \\
\Xi^-   &  \Xi^0 & - \sqrt{\frac23}\Lambda
    \end{array}  \right] \,,
\label{matrixb}
\ee
while the quark mass matrix is
\be
m_q = \left( \begin{array}{ccc}
              \,\,\,0\,\,\, & \,\,\,0\,\,\, & \,\,\,0\,\,\, \\
              \,\,\,0\,\,\, & \,\,\,0\,\,\, & \,\,\,0\,\,\,  \\
              \,\,\,0\,\,\, & \,\,\,0\,\,\, & \,\,\,m_s\,\,\, \\
             \end{array} \right) \,,
\ee
in the approximation $m_u \simeq m_d \simeq 0 $. In eq.~ (\ref{kapnel}) we
have kept only the leading nonrelativistic contributions ( except in
the baryon kinetic energy ) and scaled
the baryon fields by $B \rightarrow B \exp(-imt)$ where $m$ is the common
$SU(3)_L \times SU(3)_R$ symmetric baryon mass.   Nuclear interactions will be
included later in this section.

The Lagrange density (\ref{kapnel}) is obtained in chiral perturbation theory
as an expansion in powers of $\Lambda^{-1}$, where,  $\Lambda \simeq 1\,$GeV
is the chiral symmetry breaking scale~\cite{georgi,mangeo}.  The
operators in the expansion are invariant under $SU(3)_L \times SU(3)_R$. In
writing eq.\ (\ref{kapnel}), terms of higher order in $\del/\Lambda$ or
$m_q/\Lambda$ and also those which contain four or more baryon fields are
ignored. The expansion in terms of $\del_0/\Lambda$ is expected to be reliable,
since the time dependence of the condensate is $\exp(-i\mu t)$ ~\cite{baym} and
in neutron stars,  the chemical potential $\mu \simeq m_\pi$.  In analyzing
kaon--nucleon scattering, the expansion parameter, $m_K/\Lambda$, is not small
and higher order terms in the derivative expansion must be
included~\cite{kapnel,BLRT}.
Furthermore, spatial derivatives vanish in this case,
since we only consider  s--wave condensation.

The terms appearing in eq.~(\ref{kapnel}) are as follows. The first is the
kinetic energy of the nonlinear sigma field. We take $f$ to be the pion decay
constant $f=93$ MeV.  The second term is the meson mass term, with $m_K^2 = 2
c m_s / f^2$. The third term is the kinetic energy of the baryons.   In the
$\pi N$ sector, the fourth term is simply the s--wave interaction in the
nonlinear Weinberg Lagrangian~\cite{weinberg}.  The fifth and sixth terms
together reduce to the $\pi N$ p--wave interaction in the same model. The
constants  $F=0.44$ and $D=0.81$ are given by weak nucleon and semileptonic
hyperon decay. Note that $D+F=g_A$, where $g_A$ is the axial vector coupling
constant. The three remaining hadronic interaction terms provide the splitting
of the masses in the baryon octet.  The Lagrange densities $\L_e$ and $\L_\mu$
are those of (massless) electrons and muons, respectively. Note that except for
the baryon octet splitting terms, the Lagrange density (\ref{kapnel}) in the
$\pi N$ sector is equivalent to that employed earlier by many authors in the
study of pion condensation in neutron star
matter~\cite{baymflow,au,baymcamp,weisebrown,bcdm}.

Kaon--nucleon scattering
data does not sufficiently constrain the parameters of the model in its present
form, since higher orders in the derivative expansion must be included in the
analysis of scattering data.
The constants{\footnote{The values $a_1m_s=-67$ MeV and $a_2m_s=125$ MeV give a
better overall fit to the masses of the strange baryons; here we employ the
larger $a_2m_s$ that fits the $\Lambda$ mass accurately as in
ref.~\cite{polwise}.  This choice has very little influence on the results. }}
\be
a_1m_s =-67~~{\rm MeV}\qquad {\rm and} \qquad  a_2m_s=134~~{\rm MeV}
\ee
are determined from the relations
\be
m_\Sigma - m_N &=& 2a_2m_s \,,\nonumber \\
m_\Lambda-m_N &=& (2/3)(a_2-2a_1)m_s \,,\nonumber \\
m_\Xi - m_N &=& 2(a_2-a_1)m_s \,.
\ee
\noindent
There is a large uncertainty associated with the coefficient $a_3m_s$, which
is related to the (poorly known) value of the strangeness content of the
proton $\langle \bar{s}s \rangle_p $, and to the kaon--nucleon sigma term
$\Sigma^{K N}$~\cite{kapnel,polwise,jafkor}.
The sigma term may be related to meson--nucleon forward scattering amplitudes,
but the analysis involves a delicate extrapolation to unphysical momentum
transfer.
Depending on the method of analysis used,
the strangeness content of the proton ranges from $\langle
\bar{s}s \rangle_p = 0$ to the larger value of
$\langle \bar{s}s \rangle _p = 0.21 \langle \bar{u}u +  \bar{d}d +  \bar{s}s
\rangle _p $~\cite{donnap}.
{}From  eq.~(\ref{kapnel}) we obtain
\be
m_s \frac{\del m_p}{\del m_s} = - 2(a_2 + a_3) m_s\,,
\ee
which from QCD, equals $m_s \langle \bar{s} s \rangle_p$.
Here, we employ the two limiting values
$a_3m_s=-134$ MeV, corresponding to no strangeness, and $a_3m_s = -310$ MeV,
corresponding to strangeness of the order of 20\% in the
proton~\cite{polwise}.
The sigma term is given by
\be
\Sigma^{K N} = - \frac 12 ( a_1 + 2a_2 + 4a_3 ) m_s \,.
\label{sigmaterm}
\ee
\noindent
The possible values of the sigma term that we consider are therefore between
$\Sigma^{K N} = 168$ MeV and $\Sigma^{K N} = 520 $ MeV.
The kaon--nucleon sigma term provides the attractive interaction between
nucleons and kaons that
drives the kaon effective mass down with increasing density and
leads to the condensation of kaons.
The magnitude of $\Sigma^{K N}$ is influential in determining
the threshold and magnitude of the condensation.

The threshold density for kaon condensation is the density at which
the total energy density of matter is lowered by the introduction of a
kaon condensate.  The most tractable assumption is that the
transition to a state containing a condensate occurs for an
arbitrarily small amplitude of the condensate.  The condensation
threshold, in this case, is given by the dependence of the energy
density on small values of the kaon field amplitude only.  However, our
knowledge of loop effects contributing to the energy density is
incomplete. In addition, at the mean field level, the Lagrange density
(\ref{kapnel}) involves nonlinear interactions among the meson and
baryon fields.  In principle, either could conspire to produce a
threshold for which the energy density would be lowered by introducing
a condensate of {\it finite} amplitude, implying that the phase
transition is first order.  When we study the full nonlinear $K N$
interactions at the mean field level, we shall see that
the transition indeed occurs for small values of the condensate
amplitude.  We assume that the inclusion of loop
effects does not alter this conclusion drastically.

If only s--wave interactions are retained, the condensate is spatially uniform.
The meson condensate is then characterized by the time dependence
(see Baym~\cite{baym})
\be
\langle K^- \rangle = v_K e^{-i\mu t} \,\,
,\,\, \langle K^+\rangle = v_K e^{i\mu t}.
\label{condensate}
\ee
Here the amplitude $v_K$ is taken to be real. The non--linear interactions are
conveniently parametrized in terms of  the rotation angle $\theta$
\be
\theta = \frac{\sqrt{2}v_k}{f} \,,
\ee
which is analogous to the conventional rotation angle of the  $\sigma$--model
(see, for example, ~\cite{bkr,aubaym}).    In the sector involving only kaons
and nucleons, and excluding for the moment nucleon--nucleon
interactions, eq.\ (\ref{kapnel}) reads
\be
{\cal L}
&=& f^2  \frac{\mu^2}{2} \sin^2\theta
- 2 m_K^2 f^2 \sin^2 \frac{\theta}{2}  \nonumber\\
&+& \overline{n} i\partial_{\mu} \gamma^\mu n
+   \overline{p} i\partial_\mu \gamma^\mu p \nonumber\\
&-&  n^{\dagger}n
\left( -\mu \sin^2 \frac{\theta}{2}
+ ( 2 a_2 + 4 a_3 ) m_s \sin^2 \frac{\theta}{2} \right) \nonumber\\
&-&  p^{\dagger}p
\left( - 2 \mu \sin^2 \frac{\theta}{2}
+ ( 2a_1 + 2a_2 + 4a_3 ) m_s \sin^2 \frac{\theta}{2} \right)  \,.
\label{knexp}
\ee
In eq.\ (\ref{knexp}), we have absorbed the constant $(2a_2+2a_3)m_s$
into the nucleon mass.

Note that the kaon--nucleon interaction induced by the   sigma term $\Sigma^{K
N}$ and the Weinberg s--wave interaction are both  attractive in the s--wave
channel.  Also note the sign and magnitude of the $K N$ Weinberg s--wave
interaction in relation to that of the $\pi N$ interaction.  For neutrons, the
$KN$ interaction is attractive and equal in magnitude to the repulsive $\pi N$
interaction. For protons, both are attractive but the $KN$ interaction is twice
as strong as the $\pi N$ interaction.   These relations are given by the
virtual $\rho$ and $\omega$ vector meson exchange mediating the meson--nucleon
interactions from which the interactions under consideration are obtained in
the zero range limit. 

\newpage
\bigskip
{\ni {2.2. EQUILIBRIUM CONDITIONS }}
\bigskip
\bigskip

We first wish to find the ground state of the system of kaons,
nucleons, electrons and muons in the presence of a condensate. The
ground state in the absence of condensation will emerge naturally as a
limiting case of the ground state with a nonvanishing condensate. The
energy must be minimized, simultaneously satisfying local charge
neutrality and chemical equilibrium.  The chemical equilibrium
conditions are enumerated below.

For the densities to be considered, both $\beta$--decay and inverse
$\beta$--decay
\be
n \rightarrow p + e^- + \overline{\nu_e} \,\,\,,\,\,\,
p + e^- \rightarrow n + \nu_e
\label{ebeta}
\ee
take place. In cold neutron star matter, we may also assume that neutrinos
generated by the reactions (\ref{ebeta}) leave the system freely. This implies
\be
\mu_n - \mu_p = \mu_e \,,
\label{ebeta2}
\ee
where $\mu_n$, $\mu_p$ and $\mu_e$ are the chemical potentials (Fermi energies)
of neutrons,  protons and electrons, respectively. (In sect. 5, we
will also consider the case in which neutrinos are trapped in matter,
which leads instead to $\mu_n-\mu_p=\mu_e-\mu_{\nu_e}$).  In addition, when
the electron Fermi energy is large enough (i.\ e.\ greater than the muon mass),
it is energetically favorable for the electrons to convert to muons
\be
e^- \rightarrow \mu^- + \overline{\nu}_\mu + \nu_e \,.
\label{mubeta}
\ee
This occurs roughly at nuclear matter density, $\rho_0 = 0.16 ~{\rm fm}^{-3}$.
Denoting the muon chemical potential by $\mu_\mu$, the chemical equilibrium
established by the process (\ref{mubeta}) and its  inverse is given by
\be
\mu_\mu = \mu_e
\label{mubeta2}
\ee
If the kaon effective mass is sufficiently low in dense matter, the following
strangeness changing processes can occur:
\be
n&\leftrightarrow& p+K^-,\label{p1} \nonumber \\
e^- &\leftrightarrow& K^- +\nu_e\label{p2}
\label{strch}
\ee
Assuming that these processes take place fast
enough to establish chemical equilibrium, we have
\be
\mu_n - \mu_p = \mu \,\,\,,\,\,\, \mu_e = \mu
\label{sc2}
\ee
where $\mu$ is the chemical potential of the $K^-$ condensate. (The chemical
potential of the $K^+$ condensate is $-\mu$).

{}From   eqs. (\ref{ebeta2}),
(\ref{mubeta2}) and (\ref{sc2}),  we see that $\mu$ is the overall charge
chemical potential. Charge neutrality can be implemented by minimizing
\be
{\tilde{\cal H}}=
{\cal H} + \mu (\rho_p - \rho_K - \rho_e - \rho_\mu ) \,,
\label{htil}
\ee
where ${\cal H}$ is the Hamiltonian density,
derived from the Lagrange density in the usual manner, and $\rho_p$,
$\rho_e$ and $\rho_\mu$ are the number densities of protons, electrons
and muons, respectively.  The net negative charge in the kaon fields is
\be
\rho_K = \rho_{K^-} - \rho_{K^+}\,, \label{kcharge}
\ee
where $\rho_{K^-}$ and $\rho_{K^+}$ are the number densities of
negatively and positively charged kaons, respectively.   For the Lagrangian
eq.~(\ref{kapnel}), $\rho_K$ is determined from the definition
\be
\rho_K = i ( K^+ p_{K^+} - K^- p_{K^-} )
\label{rhok}
\ee
where the $p_{K^\pm}$ is the momentum conjugate to the
$K^\pm$ field and is given by
\be
p_{K^+} = \frac{ \del \L }{ \del \dot{K}^+ }; \qquad
p_{K-}= p_{K^+} ^ \dagger \,.
\label{pk}
\ee
Since the interaction Lagrangian contains
time derivatives the canonical momenta are not the free field expressions.
The Hamiltonian density is
\be
{\cal H} =  i n^{\dagger} \dot{ n } + i p^{\dagger} \dot{ p }
+ p_{K^-} \dot{K}^-  + p_{K^+} \dot{K}^+ - {\cal L} \,.
\ee
Using the non--relativistic approximation for the nucleons,
\be
{\cal H} &=&
- f^2  \frac{\mu^2}{2} \sin^2\theta
+ 2 m_K^2 f^2 \sin^2 \frac{\theta}{2}  \nonumber\\
&+&  \frac{1}{2m} \nabla n ^{\dagger} \cdot \nabla n
 +   \frac{1}{2m} \nabla p ^{\dagger} \cdot \nabla p \nonumber\\
&+&  n^{\dagger}n
\left( -\mu \sin^2 \frac{\theta}{2}
+ ( 2 a_2 + 4 a_3 ) m_s \sin^2 \frac{\theta}{2} \right) \nonumber\\
&+&  p^{\dagger}p
\left( - 2 \mu \sin^2 \frac{\theta}{2}
+ ( 2a_1 + 2a_2 + 4a_3 ) m_s \sin^2 \frac{\theta}{2} \right)  \nonumber\\
&+& \mu \rho_K + {\cal H}_e + {\cal H}_\mu \,,
\ee
where $\rho_K$ is given in eq.~(\ref{kcharge}). Thus
eq.~(\ref{htil}) takes the form
\be
\tilde{{\cal H}}  ( \rho_n, \rho_p, \mu, \theta  )
&=&
- f^2  \frac{\mu^2}{2} \sin^2\theta
+ 2 m_K^2 f^2 \sin^2 \frac{\theta}{2}  \nonumber\\
&+&  \frac{1}{2m} \nabla n ^{\dagger} \cdot \nabla n
 +   \frac{1}{2m} \nabla p ^{\dagger} \cdot \nabla p \nonumber\\
&+&  n^{\dagger}n
\left( -\mu \sin^2 \frac{\theta}{2}
+ ( 2 a_2 + 4 a_3 ) m_s \sin^2 \frac{\theta}{2} \right) \nonumber\\
&+&  p^{\dagger}p
\left( \mu
- 2 \mu \sin^2 \frac{\theta}{2}
+ ( 2a_1 + 2a_2 + 4a_3 ) m_s \sin^2 \frac{\theta}{2} \right)  \nonumber\\
&+& {\cal H}_e - \mu \rho_e
\,\,\,+\,\,\, {\cal H}_\mu - \mu \rho_\mu
\label{htilde}
\ee

The nucleon single particle energies are given by
$\epsilon_i = \delta \tilde{{\cal H}}/\delta \rho_i$, where $i=(n,p)$.
Expanding the nucleon wave functions in eq.\ (\ref{htilde})  in a plane wave
basis,  the result is
\be
\epsilon_n &=&
\frac{p^2}{2m}
 -\mu \sin^2 \frac{\theta}{2}
+ ( 2 a_2 + 4 a_3 ) m_s \sin^2 \frac{\theta}{2} \nonumber\\
\epsilon_p &=&
\frac{p^2}{2m}
+  \mu
- 2 \mu \sin^2 \frac{\theta}{2}
+ ( 2a_1 + 2a_2 + 4a_3 ) m_s \sin^2 \frac{\theta}{2}
\label{singlep}
\ee
Note that due to the absence of p--wave interactions there is no mixing among
the neutron and proton wave functions.

To construct a state of fixed baryon number, we occupy single particle
momentum states up to momenta $p_{F_n}$ for neutrons and $p_{F_p}$ for protons.
This leads to the energy density $\tilde{\epsilon}$:
\be
\tilde{\epsilon} ( \rho_n, \rho_p, \mu, \theta  ) &=&
\frac{3}{5}\rho_n\frac{(3\pi^2\rho_n)^{\frac{2}{3}}}{2m}
+\frac{3}{5}\rho_p\frac{(3\pi^2\rho_p)^{\frac{2}{3}}}{2m} \nonumber\\
&-& f^2  \frac{\mu^2}{2} \sin^2\theta
+ 2 m_K^2 f^2 \sin^2 \frac{\theta}{2}  \nonumber\\
&+&  \mu\rho_p - \mu ( \rho_n + 2 \rho_p ) \sin^2 \frac{\theta}{2} \nonumber\\
&+& ( \rho_p 2a_1 + \rho 2a_2 + \rho 4a_3 )
m_s \sin^2\frac{\theta}{2} \nonumber \\
&+& \tilde{\epsilon}_e
\,\,\,+\,\,\, \eta(|\mu|-m_\mu) \tilde{\epsilon}_\mu,
\label{edens}
\ee
where $\rho=\rho_n + \rho_p$ is the nucleon density, and $\eta(x)$ is the
Heaviside function ($\eta(x)=1$ if $x>0$ and $\eta(x)=0$ if $x<0$).  The
contributions of  the filled Fermi seas of the leptons are
\be
\tilde{\epsilon}_e =
- \frac{\mu_e^4}{12\pi^2}
\label{etildee}
\ee
and
\be
\tilde{\epsilon}_\mu
&=& \epsilon_\mu - \mu \rho_\mu \nonumber\\
&=& \frac{m_\mu^4}{8\pi^2}
\left\{ (2t^2+1)t \sqrt{t^2+1} - \ln ( t^2 + \sqrt{t^2+1} ) \right\}
- \mu \frac{p_{F_\mu}^3}{3 \pi^2}
\ee
where $p_{F_\mu} = \sqrt{ \mu^2 - m_\mu^2 }$ is the muon Fermi momentum
and $t = p_{F_\mu} / m_\mu $. We have allowed for the possibility that
$\mu<0$, which indicates an excess of positrons over electrons,
a situation that could occur at high densities when a condensate is present.

Up to this point, we have ignored interactions among nucleons. We include them
in a minimal fashion by adding the energy density arising from nuclear
interactions to that given by eq.\ (\ref{edens}). We use the simple
parametrizations of Prakash, Ainsworth and Lattimer~\cite{pal} which reproduce
the nuclear matter energy of more realistic microscopic calculations.  The
resulting energy density is conveniently expressed in terms of the proton
fraction $x$ and the nucleon density ratio $u=\rho/\rho_0$, where
$\rho_0 = 0.16~ {\rm fm}^{-3}$ is the equilibrium nuclear matter density.  The
proton and neutron densities, $\rho_p$ and $\rho_n$, are given by
\be
\rho_p = x\rho \,\,\,\,,\,\,\,\, \rho_n = (1-x)\rho
\,\,\,\,,\,\,\,\, \rho = u\rho_0
\ee
Including nuclear interactions, eq.~(\ref{edens}) is replaced by
\be
\tilde{\epsilon}( u, x, \mu, \theta  )
&=& \frac{3}{5} E_F^{(0)} u^{\frac{5}{3}} \rho_0 \nonumber\\
&+& V(u) + u\rho_0(1-2x)^2S(u) \nonumber\\
&-& f^2  \frac{\mu^2}{2} \sin^2\theta
+ 2 m_K^2 f^2 \sin^2 \frac{\theta}{2}  \nonumber\\
&+& \mu u \rho_0 x -   \mu u\rho_0 (1+x) \sin^2 \frac{\theta}{2} \nonumber\\
&+& ( 2a_1x + 2a_2 + 4a_3 )
m_s u\rho_0  \sin^2\frac{\theta}{2} \nonumber \\
&+& \tilde{\epsilon}_e
\,\,\,+\,\,\, \eta(|\mu|-m_\mu) \tilde{\epsilon}_\mu,
\label{epst}
\ee
where $E_F^{(0)} = (p_F^{(0)})^2/2m$ and $p_F^{(0)} =
(3\pi^2\rho_0/2)^ {\frac{1}{3}}$ are the Fermi energy and momentum,
respectively, at nuclear saturation density. In eq.\ (\ref{epst}),
$V(u)$ is the potential contribution to the energy density of
symmetric nuclear matter.  Its explicit form is not necessary for the
evaluation of the threshold density for kaon condensation, or for the
evaluation of the condensate amplitude beyond threshold.  It is,
however, needed for the construction of the equation of state (the
pressure--density relation).  Following ref.~\cite{pal},
\be
V(u) &=& \frac 12 A u^2 \rho_0
+ \frac { B u^{\sigma+1} \rho_0 }{ 1 + B'u^{\sigma - 1} } \nonumber\\
&+& 3 u \rho_0 \sum_{i = 1,2}
C_i \left( \frac {\Lambda_i} { p_F^{(0)} }\right)^3
\left( \frac{p_F}{\Lambda_i}  - \arctan \frac{p_F}{\Lambda_i} \right)
\label{vu}
\ee
Here $p_F$ is the Fermi momentum, related to $p_F^{(0)}$ by
$p_F=p_F^{(0)}u^{1/3}$. The parameters $\Lambda_1$ and $\Lambda_2$ parametrize
the finite--range forces between nucleons. The parameters $A$, $B$, $\sigma$,
$C_1$, $C_2$ and $B'$, a small parameter introduced to maintain causality, are
determined from constraints provided by properties of nuclear matter at
saturation~\cite{pal}.  To explore the dependence on the stiffness of the EOS,
the nuclear incompressibility is parametrically varied.    Numerical values of
the parameters are given in table~2.

The energy density of asymmetric nuclear matter
may be written as an expansion in $(1-2x)^2$ about the symmetric
matter energy density. To a very good approximation, it is sufficient
to retain in this expansion only the quadratic term as given
above~\cite{pal}.  The nuclear symmetry energy $S(u)$ is then given by
the sum of kinetic and potential terms:
\be
S(u) &=& \left( 2^{\frac{2}{3}}-1 \right)
\frac{3}{5}E_F^{(0)}\left( u^{\frac{2}{3}} - F(u) \right)
     + S_0 F(u)
\ee
where $S_0 \simeq 30~{\rm MeV}$ is the bulk symmetry energy parameter.
The function $F(u)$ parametrizes the potential contribution to the
symmetry energy and satisfies $F(0)=0$ and $F(1) = 1$.  Three
representative forms for $F(u)$ considered in ref.~\cite{pal} and in
this work are
\be
F(u) = u, \qquad F(u) = \frac{2u^2}{1+u} \qquad {\rm and} \qquad
F(u) = \sqrt{u} \label{fu}
\ee
The functions (\ref{fu}) mimic the results of more realistic models.
Inclusion of the nuclear symmetry energy, which has
heretofore not generally been considered in the context of meson
condensates, is important in obtaining the correct matter composition,
and affects both the threshold and amplitude of the kaon condensate.
The symmetry energy favors the addition of protons to nuclear matter
and contributes to the large proton fraction $x$ found in the ground
state in the presence of a meson condensate.

In order to determine the energy density of the charge neutral ground state,
$\epsilon$, we find the ground state of ${\tilde{\cal H}}$ for fixed
baryon density.  Denoting the energy of the ground state of ${\tilde{\cal H}}$
by ${\tilde \epsilon}$, this is done by extremizing $\tilde\epsilon$
with respect to $x, \mu,$ and $\theta$:
\be
\frac{\del {\tilde \epsilon}}{\del x} = 0; \quad
\frac{\del {\tilde \epsilon}}{\del \mu} = 0; \quad
\frac{\del {\tilde \epsilon}}{\del \theta} = 0,
\label{charneut}
\ee
which leads to the following set of equations:
\be
\mu = 4(1-2x)S(u)\sec^2 \frac{\theta}{2}
- 2a_1 m_s \tan^2\frac{\theta}{2} \,,
\label{betatheta}
\ee
\be
f^2\mu \sin^2\theta + u\rho_0 (1+x)\sin^2\frac{\theta}{2} - xu\rho_0
 + \frac{ \mu^3 }{ 3\pi^2 }
 + \eta(|\mu|-m_\mu) \frac{ (\mu^2-m_\mu^2)^{3/2}}{ 3\pi^2 }
= 0, \nonumber \\
\label{cntheta}
\ee
and
\be
\theta &=& \cos^{-1}\left[ \frac {1}{\mu^2}
\left( m_K^2 - \frac{\mu}{2f^2} u \rho_0 ( 1 + x )
+ \frac{ u \rho_0 }{ 2f^2 } ( 2a_1x + 2a_2 + 4a_3 )m_s \right)\right]
\nonumber\\
\label{thetamin}
\ee
The first equation shows how the charge chemical potential varies from
the value imposed by the nucleons when a condensate forms.  The second
equation is simply a statement of charge neutrality. The third
equation determines the amplitude of the condensate.  Thus, from
eq.~(\ref{cntheta}) we may
read off the negative charge density in the condensate $\rho_K$ as
\be
\rho_K =
f^2\mu \sin^2\theta + u\rho_0 (1+x)\sin^2\frac{\theta}{2} \label{ncharge}
\ee

\newpage
\bigskip
{\ni {2.3. NUMERICAL RESULTS }}
\bigskip
\bigskip

The critical density may be found from the above relations simply by setting
$\theta=0$.  Although the result cannot be expressed in closed form, a numerical
determination of the critical density is straightforward.
Critical densities from eq.~(\ref{thetamin}) are displayed in table~1.
The choice of $a_3m_s$ has a greater effect on the critical
density than the choice of $F(u)$ or $S_0$.  In this model, the upper and lower
bounds for the critical density are $u_c=4.95$ and $u_c=2.30$, respectively. The
inclusion of muons in beta equilibrium increases the critical density slightly
over that obtained when muons are ignored. For example, the critical density in
the first case in table~1
is $u_c=3.90$ when only electrons are
included. This behaviour is expected, since the electron chemical potential in
beta equilibrium is higher if muons are ignored and the threshold condition is
therefore fulfilled at a lower density.

An equivalent method to determine the threshold density for kaon condensation
is to determine the density at which the charge chemical potential $\mu_0$
{\it in the absence of a condensate} corresponds to the energy of a physical
particle.  This corresponds to solving $D^{-1}(\mu_0)=0$, where $D^{-1}(\omega)
=  \omega^2 - m_K^2 - \Pi (\omega)$ is the inverse kaon propagator in medium,
and  $\Pi (\omega)$ is the kaon self--energy in matter.  (See appendix A for
more details.) In fig.~1,
we show the energy of zero--momentum physical particles,
eq.\ (\ref{omega}).   Superimposed is the charge chemical potential in the
absence of a condensate, for the maximum and minimum values of $a_3m_s$
considered in this work.    The point of intersection of the energy
$\omega^-(u)$ with $\mu_0(u)$  is $u_c$ (see table~1).
Note that for $a_3m_s=-310$ MeV, the kaon energy and threshold are lower than
for $a_3m_s=-134$ MeV.  This may be rephrased in terms of the kaon--nucleon
sigma term $\Sigma^{KN}$ (see sect. 2.\ 1). The larger the $\Sigma^{KN}$, the
lower the kaon energy and condensation threshold.

In tables 3, 4 and 5,
we show  the angle
$\theta$ at which the energy is minimized as a function of density.
Also tabulated are the chemical potential $\mu$, the gain in energy density
$\Delta \epsilon = \epsilon (u,\theta)-\epsilon(u,0)$ (due
to the appearence of the condensate), the proton
fraction $x$, the kaon fraction $x_K \equiv  \rho_K/\rho$, the electron
fraction $x_e \equiv  \rho_e / \rho$ and the muon fraction $x_\mu \equiv
\rho_\mu / \rho$ for the minimum energy configuration. {}From charge neutrality
$x = x_K + x_e + x_\mu $. In tables 3 through 5,
$a_3m_s=-134 $ MeV, $-222$ MeV and $-310$ MeV, respectively.

The results in tables 3 through 5
indicate that the transition from normal matter to matter with a condensate is
of second order, with the amplitude of the condensate $v_K$ or $\theta$ serving
as an order parameter. Following Migdal~\cite{migdal}, we consider the
transition in a Ginzburg--Landau description by expanding the energy density
eq.\ (\ref{epst}) to second order in $\theta^2$.    We then find that just
above threshold the energy density has the density dependence
\be
\Delta \epsilon ( u )= - \frac 12 \beta ( u - u_c )^2 \,,
\ee
where $\beta$ is a positive constant independent of density. In the case of pion
condensation, it was shown by Dyugayev~\cite{dyugayev} that in the vicinity of
the critical point, long range pion--pion interactions cause the transition to
matter with a pion condensate to be weakly first order.

Above the threshold density for condensation, the condensate contribution to
the energy density and pressure are:
\be
\epsilon_K &=& f^2  \frac{\mu^2}{2} \sin^2\theta
+ 2 m_K^2 f^2 \sin^2 \frac{\theta}{2}
+ ( 2a_1x + 2a_2 + 4a_3 ) m_s u\rho_0 \sin^2\frac{\theta}{2} \,, \nonumber \\
P_K &=& f^2 \frac {\mu^2}{2} \sin^2 \theta
     -  2 m_K^2f^2 \sin^2\frac{\theta}{2} \,.
\ee
Note that $\mu$, $x$ and $\theta$ implicitly depend on the density via the
extremization conditions in eqs.~(\ref{betatheta}) through (\ref{thetamin}).

Prior to condensation, $\mu$ is a gradually increasing function of the density
ratio $u$ with the
rate of increase controlled essentially by the density dependence of the
symmetry energy $S(u)$.  With the onset of condensation, the net negative
charge in the kaon fields causes the lepton concentrations to decrease which
reverses the behavior of $\mu$ with density.  {}From tables 3, 4 and 5,
we see that, in each case, there is a certain
value of $u$ for which $\mu = 0$.  This implies $\rho_e = 0$  and hence $\rho_K
= \rho_p$.  Several other simplications occur in the case when $\mu=0$.
Charge neutrality (\ref{cntheta}) gives the simple relation
\be
x = \tan^2 (\theta/2) \,.
\label{xreal}
\ee
Eq.~(\ref{betatheta}) leads to the relation
\be
S(u)={a_1m_sx\over 2(1-2x)(1+x)} \,,
\label{mu0}
\ee
and, from the condition that $\theta$ is real, eq.~(\ref{thetamin}) gives
\be
u\rho_0 &=& -{2m_K^2f^2\over m_s(2a_1x+2a_2+4a_3)} \nonumber \\
        &=& \frac {m_K^2f^2}{\Sigma_{KN}-(1-2x)a_1m_s/2} \,.
\label{urho0}
\ee
Eq.~(\ref{mu0}) illustrates that, since $S(u)>0$ and $a_1<0$, $x$ must be
greater than 1/2 when $\mu=0$. {}From Tables 2--4 it is clear when
$\mu=0$ that $x\simeq1/2$.  Combining eqs.~(\ref{mu0}) and
(\ref{urho0}) and setting $x_{\mu=0}=1/2+y$, where $y$ is a small
number, one finds
\be
y\simeq\frac{a_1\rho_0m_s^2(a_1+2a_2+4a_3)}{24m_K^2f^2(S(u)/u)}.
\label{yapprox}
\ee

In the case in which $F(u)\sim u$ one has $S(u)\simeq S_0u$ for $u>>1$.
Hence,
\be
y \simeq \frac {a_1\rho_0m_s^2(a_1+2a_2+4a_3)}{24S_0m_K^2f^2}
\label{xapprox}
\ee
and
\be
\theta_{\mu=0}=2\arctan{{\sqrt x_{\mu=0}}}
\simeq 2\left[ \arctan\left( \frac {1}{{\sqrt 2}}\right)
+ \frac {\sqrt 2}{3}y \right]
\ee
The density $u_{\mu = 0}$ is obtained by inserting $x_{\mu=0}$ in
eq.~(\ref{urho0}). In the case $a_3m_s=-222$ MeV, one finds $y=0.04$ and
$u_{\mu=0}=4.95$, for example.  The magnitudes of $y$ and $1/u_{\mu=0}$
increase roughly proportionately with $a_3m_s$; this behavior is followed in the
tables. Also note that $u_{\mu=0}$ is expected to be insensitive to the
symmetry energy term $S_0$, since $y\ll 1$.  In more generality, the properties
of the condensate are not very sensitive to the behavior of the symmetry energy
at high densities for the models we have chosen.  This behavior will change,
however, if the symmetry energy becomes negative, as they do in some early
potential model calculations.

When $\mu = 0$, the net negative charge in the condensate is balanced by
protons, i.e.,
\be
\rho_K/\rho_0 = x_K(\mu =0) = x \,,
\ee
which from eq.~(\ref{xapprox}) has the approximate value of $1/2$.
The condensate energy density and pressure for $\mu = 0$ are
particularly simple:
\be
\epsilon_K = 0 \qquad {\rm and} \qquad P_K &\simeq &  - (2/3) m_K^2 f^2 \,,
\ee
wherein for $P_K$ the result $\theta_{\mu=0} \simeq 2~\arctan (1/{\sqrt 2})$
has been utilized.

For higher densities, $\mu < 0$.  At these densities, the matter
contains {\rm positrons} in chemical equilibrium with the kaon
condensate and nuclear matter. The charge of the positrons and protons
is balanced by the charge of the kaon condensate, which remains
negative ($\rho_K > 0 $). The more negative the $a_3m_s$, the lower
the threshold for positron production. {}From these tables, we also see
that there is a further threshold for the production of positive muons
at somewhat higher densities. The following section will include a
discussion on whether the relevant densities are reached in realistic
neutron star calculations.

\bigskip
\bigskip
\addtocounter{section}{1}
\setcounter{equation}{0}
{\centerline {\bf 3. Neutron stars with a kaon condensate }}
\bigskip
\bigskip

In this section, we examine whether or not a kaon condensate can exist
in the core of a neutron star, and if so, how the presence of a
condensate would affect the gross features of the star, such as its
mass and radius. In general, the maximum mass of a neutron star is low
if a soft nuclear equation of state is employed.  It is therefore
important to know if equations of state with a kaon condensate, which
provides additional softening, will be stiff enough to produce a star
of mass greater than \sol, the value for the largest neutron star in
the PSR 1913+16 system and the lower limit for the maximum neutron
star mass.  We shall see that this constraint is not satisfied for all
combinations of the underlying nuclear equation of state (as given by
the parameterization of ref.~\cite{pal}, for example) and the value of
the condensate parameter $a_3m_s$, which specifies the strangeness
content of the proton.

\bigskip
\bigskip
{\ni {3.1. THE EQUATION OF STATE }}
\bigskip
\bigskip

For densities $\rho > 0.08~{\rm fm}^{-3}$,  we use the parametrization of the
nuclear matter equation of state given in ref.~\cite{pal}.   The EOS for
$\rho > \rho_c$ is that described in the previous section.
For densities $0.001< \rho < 0.08~ {\rm fm}^{-3}$, we employ the equation of
state of Negele and Vautherin~\cite{negvaut}, while for low densities ($\rho <
0.001~ {\rm fm}^{-3} $), we use the Baym--Pethick--Sutherland~\cite{bps}
equation of state.  These choices have almost no bearing on our
results.

In many cases studied, the compressibility $K = 9 dP/d\rho|_{\rho_c}$
at threshold is found to be negative.  This implies that just above
threshold, matter will be compressed until hard core repulsion results
in a positive compressibility~\cite{migdal}.  Following
Migdal~\cite{migdal}, we use the Maxwell construction to maintain a
positive compressibility. Note that the Maxwell construction is
normally applied to phase transitions that are first order by
construction, whereas the transition here is of second order (see
sect. 2.2).

In fig.~2,
we show the pressure as a function of total mass--energy density
$\epsilon$ for $K_0=240$ MeV, $F(u)=u$ and two values of $a_3m_s$.  In
both cases shown, the EOS is softened considerably due to the
condensation.  For $a_3m_s = -222$ MeV, the condensation is so strong
that a Maxwell construction is required, with the densities
$u_{min}=2.26$ and $u_{max}=5.47$ being the lower and upper densities
of the region of constant pressure.  If the magnitude of $a_3m_s$ is
further increased, the value of $u_{min}$ continues to fall.  As a
rule, we shall consider solutions with $u_{min}<1$ to be unrealistic;
this occurs for our nucleon-nucleon force when $a_3m_s>310$ MeV and
$K_0>300$ MeV (see also figure 5).

\bigskip
\bigskip
{\ni {3.2. STELLAR STRUCTURE RESULTS }}
\bigskip
\bigskip

We turn now to the solutions of the TOV equation, the equation of
general relativistic hydrostatic equilibrium. In fig.~3,
we show the mass versus the central density
$\rho_{cent}$ for $a_3m_s = -134$ MeV, $F(u)=u$ and for various values
of $K_0$.  We also show the dependence of mass on the radius for the
same cases.  As expected, the maximum mass decreases as the equation
of state becomes softer (smaller values of $K_0$).  The signature of
the Maxwell construction used for the softer equations of state is
the flat portion of the mass vs. central density curve.  In all
cases studied, regions with $dM/d\rho_{cent}<0$ do not correspond to
stable solutions.
%
In fig.~4,
we show analogous results for the parameter $a_3m_s = - 222 $ MeV.
For the case $a_3m_s = -310$ MeV, and for $K_0<300$ MeV, no solution
was found with $u_{min} \geq 1$ and this case is not considered
further in this paper.

Figure 5 shows the dependence of the maximum mass as a function of
$K_0$ and $a_3m_s$.  The choice of symmetry energy and the values of
$a_1$ and $a_2$ do not influence the contours shown in figure 5
appreciably.  It is somewhat surprising that the maximum mass is only
slightly sensitive to the value of $a_3m_s$.  Since any model for cold,
catalyzed matter must satisfy the binary pulsar constraint that
$M_{max}>1.44 M_\odot$, some combinations of $K_0$ and $a_3m_s$ are not
permitted.  Effectively, if a condensate occurs, $K_0$ must be greater
than about 180 MeV.  If a condensate does not occur, the minimum
allowed value of $K_0$ is about 115 MeV (see Table 6).

In Tables 6 through 8, we list the mass $M_{max}$, the
radius, $R$, the central density $\rho_{cent}$ and the Keplerian
rotational frequency $\Omega_K$ (see below) of maximum mass stars.  In
Table 6,
the maximum mass star properties are listed for
the EOS without kaon condensation. In Table 7,
the maximum
mass star properties are given for $a_3m_s = -134$ MeV, for various
compression modulii $K_0$, and the three choices for the potential
contribution to the symmetry energy $F(u)$.    In Table 8,
we show the maximum mass star parameters for $a_3m_s = - 222 $ MeV and
the choice $F(u)=u$.
Neutron star properties
have a much stronger dependence on $K_0$ than on the
choice of $F(u)$.
For decreasing
compression modulus, the central density of the maximum mass star
increases.  Correspondingly, the radius of the
maximum mass star decreases.  A kaon condensate with even the smallest
possible magnitude of $a_3m_s$ permitted in our model (-134 MeV) has large
consequences.  The maximum masses are considerably reduced from the
case when no condensate appears.

The structural changes in the star induced by the occurence of a condensate has
other implications as well.  High frequency pulsars, if observed,
could set severe limits on the
possible nuclear equations of state, assuming that the pulsations are
due to rotation~\cite{fip,lpmy}.  The maximum rotation rate of a neutron star
with a given equation of state is given by the Keplerian frequency $\Omega_K$,
defined as the rotational frequency at which the orbital velocity of a particle
at the equator is equal to the equatorial surface velocity.  For rotational
frequencies greater than $\Omega_K$, the neutron star will expel mass in the
equatorial region.
General relativistic calculations assuming uniform rotation are well reproduced
by the relation~\cite{fip,lpmy}
\be
\Omega_K = 7.7 \times \left( \frac {M_{max}}{M_\odot} \right)^{1/2}
\left( \frac {R_{max}}{10~{\rm km}} \right)^{-3/2} ~~{\rm s}^{-1} \,,
\ee
where the subscript ``max'' denotes properties of the maximum mass
non--rotating star of a given EOS.
This result seems to be independent of equation of state and whether
or not condensates and/or quark matter appear.  As pointed out in
ref.~\cite{lpmy}, the softening induced by kaon condensation in the
vicinity of nuclear matter density and subsequent stiffening at higher
densities permits a large rotational frequency supporting at the same
time a mass of at least \sol, as shown in tables 7 and 8.
In general, the stiffer equations of state tend to
produce larger radii and hence lower $\Omega_K$'s.

Another implication of a kaon condensate concerns the thermal
evolution of neutron stars.  Kaon condensates would increase neutrino
emissivities by a large factor compared to conventional neutrino
processes such as the modified Urca process.  As we now discuss, the
enhanced emission is due mostly to the direct nucleon Urca process as
opposed to the kaon condensate itself.  Enhanced neutrino
emission leads to more rapid cooling of neutron stars~\cite{bkrt}, and
could have significant observational consequences.

The contribution to the neutrino emissivity from a kaon condensate due to
the direct and inverse processes in eq. (\ref{strch}) is given by
\be
{\cal {L}}_\nu &=& \frac {2\times 457\pi}{20160}
\left( \frac {G}{{\sqrt 2}}\right)^2
(m_n^{*})^2\mu(kT)^6 (1+3g_A^2) \frac {\sin^2\theta}{ 4} \sin^2\theta_C
\nonumber \\
&=& 2\times 7.6 \times 10^{25}~~{\rm {erg~cm^{-3}~sec^{-1}} } \nonumber \\
&\times&\left(\frac {m_n^{*}}{m_n} \right)^2 \left(\mu\over m_\pi\right)
T_9^6 (1+3g_A^2) \frac {\sin^2\theta}{4} \sin^2\theta_C
\label{nuemiss}
\ee
where $m_n^{*}$ is the nucleon effective mass, $T$ is the temperature,
$T_9$ is the same in units of $10^9$ K, $g_A$ is the nucleon
axial--vector coupling constant and $\theta_C$ is the Cabibbo angle.
This result generalizes the earlier estimate of Brown et
al~\cite{bkpp}, who considered only small amplitudes of the
condensate.  The only difference from the earlier result is that the
condensation amplitude factor $\theta \rightarrow \sin\theta$.

The cooling induced by the kaon
condensate is substantially more rapid  than the standard modified Urca
process, but is significantly slower than that due to the direct Urca processes
in eq.~(\ref{ebeta}), as~\cite{lpph}
\be
{\cal {L}}_\nu &\simeq& {\cal {L}}_{Urca}
\frac {\sin^2\theta}{8} \tan^2\theta_C \,.
\ee
The emissivity from the direct Urca process is suppressed by a 
factor $\cos^2(\theta/2)$ in the presence of a kaon condensate~\cite{fujii}.
For the direct Urca processes to occur in matter in which the only
baryons are nucleons, momentum conservation requires that
the magnitude of the electron concentration in matter exceeds a
value~\cite{lpph}
\be
|x_e|^{1/3}\ge |(1-x)^{1/3}-x^{1/3}|.
\label{urca}
\ee
When kaons are not condensed, this reduces to a condition that $x$
exceed a value in the range $0.11-0.15$.  Because $x$ depends
sensitively on the density dependence of the symmetry energy, it is
not clear if cooling due to the nucleon direct Urca process occurs at
all~\cite{lpph}.  If hyperons are present, however, additional
channels of cooling via the hyperon direct Urca process are possible,
but not guaranteed~\cite{PPLP}.

In the presence of a kaon condensate, however, a large proton
concentration is required to satisfy charge neutrality, and the kaon
condensate will {\it automatically} trigger the more rapid direct Urca
process.  The condition, eq.~(\ref{urca}), is generally
satisfied except for a narrow range of densities around the point
where $\mu=0$ and $Y_e$ vanishes.  Because a neutron star encompasses
a range of densities, the direct Urca process is sure to occur
somewhere in the star.  The fact that the direct Urca process
may be forbidden in a narrow range of densities in a neutron star's
interior is inconsequential, and the core of the star will rapidly cool.

Another intriguing question is whether the densities encountered in
neutron stars are sufficiently large for the production of
positrons or positively charged muons as noted in sect. 2.  As eq.
(\ref{urho0}) maintains, this density is primarily determined by the
value of $a_3m_s$.  For
$a_3m_s = - 134$ MeV, positron production sets in at $u \simeq 10$,
while the central densities achieved in maximum mass stars can in fact
be larger than this (Table 7) in some cases.
For $a_3m_s= -222$ MeV,  both $e^+$ and $\mu^+$ appear at much
lower densities ($ u = 5.0 $ and $ u = 6.5 $, respectively).   The central
densities of maximum mass stars for this case are well above these
values (see Table 8).

Let us summarize the results of this section.  Modest values of the
strangeness content of the proton lead, in this model, to the
formation of kaon condensates in the cores of neutron stars with
masses in the range of those observed.  There is a strong dependence
of the maximum mass on the overall stiffness of the EOS, but the
dependence on the strength of the condensate and the choice of the
symmetry energy is relatively weak. In the model under consideration,
some combinations of stiffness and the intrinsic strangeness content
of the proton are ruled out by the observation of a \sol mass neutron
star in the binary pulsar PSR 1913+16. The maximum rotational
frequency of stars with kaon condensates can be quite large and could
be compatible with a sub--millisecond pulsar, should one ever be
observed.  The neutrino emissivity from matter which contains a kaon
condensate is substantially larger than that due to the standard
modified Urca process, not because of condensate-induced neutrino
emission, but mostly because of the occurence of the nucleon direct
Urca process.  Such stars will cool extremely rapidly.  It is even
possible that regions within the neutron star core exist which contain
positrons and positively charged muons in chemical equilibrium with
the condensate and nuclear matter.

\bigskip
\bigskip
\addtocounter{section}{1}
\setcounter{equation}{0}
{\centerline {\bf 4. Transition to quark matter }}
\bigskip
\bigskip

Depending on the pressure--density relationship, the density of matter in  the
core of a neutron star may be $\rho \sim 1-2~{\rm fm}^{-3}$ (see tables
6 through 8).  At such high
density, the energy of quark matter may well lie below that of baryonic mattter
with kaon condensates considered in the previous section.  Here, we explore the
conditions under which a phase transition to quark matter may take place.

\newpage
\bigskip
{\ni {4.1. THE EQUATION OF STATE OF QUARK MATTER}}
\bigskip
\bigskip

For the equation of state in the quark phase, we first consider the work of
Freedman and McLerran~\cite{FM} in which the thermodynamic potential of a quark
gas to fourth order in the quark--gluon coupling, $g$, was calculated.

The Fermi gas contribution to the thermodynamic potential from a quark species
 $~i~(=u,~d~{\rm or}~s)$ of mass $m_i$ and chemical potential $\mu_i$ is
\def\ef{(\mu_i^2-m_i^2)^{1/2}}
\be
\Omega^{\rm {FG}}_i = - \frac {1}{4\pi^2}
\left\{ \mu_i\mu_i^*\left(\mu_i^2-\frac 52 m_i^2\right) + \frac 32 m_i^4
\ln \left(\frac {\mu_i+\mu_i^*}{m_i} \right)  \right\}
\ee
where $\mu_i^*=\ef$.  For massless quarks (e.g. $u$ and $d$ quarks), the Fermi
gas result simplifies to $\Omega^{\rm {FG}}_i = -(1/4\pi^2) \mu_i^4$.

The two--loop or exchange contributions are given by~\cite{BC}
\be
\Omega^{\rm {ex}}_i =  \frac {1}{4\pi^2}  \frac {2\alpha_c}{\pi}
\left\{  3\left[\mu_i\mu_i^* - m_i^2 \ln \left(\frac {\mu_i+\mu_i^*}{m_i}
\right)\right]^2 - 2(\mu_i^2 - m_i^2)^2 \right\}
\label{omex}
\ee
which reduces to $\Omega^{\rm {ex}}_i = (1/4\pi^2) (2\alpha_c/\pi) \mu_i^4$,
for massless quarks. Here, $\alpha_c \equiv  g^2/(4\pi)$.  Higher order
correlation corrections (from ring diagrams, etc. ) have only been evaluated
for the massless case, for  which
\be
\Omega^{\rm {corr}}_i = \frac {1}{4\pi^2}  \mu_i^4
\left[ (\alpha_c/\pi)^2 \ln (\alpha_c/4\pi) + 31.1~(\alpha_c/4\pi)^2 \right]
\ee
To the same order, a further contribution associated with the interference
between up and down quarks also exists, and its (lengthy) expression may be
found in ref.~\cite{FM}.  The screened charge $\alpha_c$ falls off with  the
chemical potential $\mu$ of the quark according to  the Gell--Mann--Low
equation.  To the order considered above, the  precise way in which this fall
off occurs is described in ref.~\cite{FM}.

The production of electrons by the beta decays
\be
d  & \leftrightarrow & u + e^- + {\overline \nu_e} \nonumber \\
s  & \leftrightarrow & u + e^- + {\overline \nu_e} \label{qbeta}
\ee
leads to the additional contribution
\be
\Omega_e \cong -\frac {1}{3} \frac {1}{4\pi^2} \mu_e^4
\ee
to the thermodynamic potential (the electromagnetic interactions of the
electrons give negligible contributions), where, as before, the neutrinos are
taken to have left the system.  The beta--equilibrium and charge neutrality
conditions are
\be
\mu_d = \mu_u + \mu_e = \mu_s
\ee
and
\be
Q = e\left(\frac 23\rho_u - \frac 13\rho_d - \frac 13 \rho_s - \rho_e\right) = 0
\label{qcharge}
\ee
respectively. In the above, the number densities are given by $\rho_i = -
(\partial /\partial \mu_i)\Omega$, with the baryon number density
\be
\rho = (\rho_u+\rho_d+\rho_s)/3\,. \label{bnumber}
\ee
The existence of muons is contingent on the electron Fermi energies exceeding
the muon mass.  However, as is clear from eq.~(\ref{qcharge}), quarks, in equal
amounts of each flavor, maintain charge neutrality amongst themselves, needing
the aid of electrons only when the strange quarks are massive.  Thus, electron
concentrations in quark matter are rather small, so small in fact as to
preclude the existence of muons.

The total pressure and energy density of the electro--neutral quark matter
system are thus
\be
P &=& -\Omega = - (\Omega_u + \Omega_d + \Omega_s + \Omega_{int} + \Omega_e)
\nonumber \\
{\cal E} &=& \Omega + \sum_i \mu_i\rho_i
\label{qmeos}
\ee

Fahri and Jaffe~\cite{FJ} retain terms up to order $\alpha_c$
in the  the thermodynamic potential of
eq.~(\ref{qmeos}) and choose a renormalization procedure in which $\alpha_c$ is
held fixed.  The renormalization point, $\rho_R$, for the massive $s$ quark is
chosen at $\rho_R= 313~{\rm {MeV}}$ to minimize its (unphysical) impact on
physical variables.  In this scheme, the exchange contribution from the strange
quarks is given by
\def\efs{(\mu_s^2-m_s^2)^{1/2}}
\be
\Omega^{\rm {ex}}_s &=& \frac {1}{4\pi^2}  \frac {2\alpha_c}{\pi}
\left\{  3\left[\mu_s\mu_s^* - m_s^2 \ln \left(\frac {\mu_s+\mu_s^*}{\mu_s}\right)
\right]^2 - 2(\mu_s^2 - m_s^2)^2 \right. \nonumber \\
&-& \left. 3 m_s^4 \ln^2 \frac {m_s}{\mu_s}
+ 6 \ln \frac {\rho_R}{\mu_s} \left[\mu_s\mu_s^*m_s^2 - m_s^4 \ln
\left( \frac {\mu_s+\mu_s^*} {m_s}\right) \right] \right\}
\ee
where $\mu_s^*=\efs$.

To complete the description of the quark matter phase, the results in
eq.~(\ref{qmeos}) are supplemented with the dynamics of the MIT bag by
adding a positive constant contribution, $B$, to the energy density ${\cal E}$,
whence
\be
\Omega \rightarrow \Omega + B\,, \qquad
P \rightarrow P - B\,, \qquad
{\rm and}~ \qquad
{\cal E} \rightarrow {\cal E} + B.
\ee
The constant $B$ has the simple interpretation as the thermodynamic potential
of the vacuum, and in comparisons of energy and pressure of quark--matter to
nuclear matter calculations, is regarded as a phenomenological parameter.

\bigskip
\bigskip
{\ni {4.2. NUMERICAL RESULTS }}
\bigskip
\bigskip

The transition density $\rho_t$ above which the quark phase has a
lower energy than the hadronic phase follows from the Gibbs criterion
that the chemical potentials and pressures in the two phases are equal
at $\rho_t$.  For a given EOS on the hadronic side, the magnitude of
$\rho_t$ depends sensitively on whether the hadronic EOS is soft or
stiff at the relevant densities~\cite{lpmy,PBP}.  For stiff hadronic
EOS's, $\rho_t$ is often larger than the central density of the
maximum mass star, $\rho_c$.  Soft hadronic EOS's allow matter to
exist at much higher densities, which could permit quark cores to
exist.  In addition, the transition density will depend on the bag
constant, $B$ (recall that $B$ is not well constrained in perturbative
QCD calculations of quark matter).  fig.~6 illustrates this for two
hadronic EOS's of ref.~\cite{pal} termed PAL21 ($K_0=180$ MeV and
$F(u)=u$) and PAL31 ($K_0=240$ MeV and $F(u)=u$).  The EOS with kaon
condensation refers to that with $a_3m_s=-134$ MeV in both cases. The
quark matter EOS shown corresponds to that of ref.~\cite{FM} with
$m_s=200$ MeV (as required by the fits of Gasser and
Leutwyler~\cite{GL}) and the renormalization point $\mu_0=300$ MeV and
$\alpha_c (\mu_0)=1$ (see ref.~\cite{FM} for details).  The screened
charge $\alpha_c$ decreases rapidly from $\sim 0.16$ at
$2\times\rho_0$ to $\sim 0.1$ at $\rho \geq 1~{\rm fm}^{-3}$.  The
transition density, $\rho_t$, for pure hadronic EOSs are in excess of
$\rho \cong (4-6)~\rho_0$.  A general consequence of the additional
softening provided by the kaon condensate is that the transition to
quark matter is postponed to higher densities, if it occurs at all.

Similar conclusions are also reached for the quark matter EOS (up to order
$\alpha_c$) of ref.~\cite{FJ}.  For the QM curve shown in fig.~7, $m_s=200$ MeV
$B=100~{\rm MeV/fm^{3}}$ and $\alpha_c =0.55$ independent of density.
Note that the transition to a quark phase occurs only with
small values of  $\alpha_c$.  Larger values as used in the MIT bag fits to
hadronic  spectroscopy preclude a phase transition to the quark phase.
Clearly, a transition may be achieved only through the use of correlated values
of $B$ and $\alpha_c$.

The above perturbative results for the quark phase are expected to be  reliable
only for $\rho \gg \rho_0$.   If, in a two phase description, $\rho_t$ turns
out to be only a few times $\rho_0$, firm statements about a quark core cannot
be made since the quark phase EOS is expected to receive substantial
corrections from non--perturbative effects at such densities.   Anticipating
such corrections to be significant, Bethe, Brown and Cooperstein~\cite{BBC} have
argued that such effects may be built in phenomenologically via the order
$\alpha_c$ results in eq.~(\ref{qmeos}) but taking the density dependence of
$\alpha_c$ as  $\alpha_c^{(0)}~(\rho_0/\rho)^{1/3}$.   Results from this scheme
are shown in fig.~8 for the case $m_s=200$ MeV, where it is again seen that a
transition to the quark phase is possible only if small values of $\alpha_c$
are used in the description of the quark phase.   When it occurs, the
transition takes place at densities substantially larger than $(4-5)~\rho_0$,
and only for select choices of the bag constant $B$.  As with the other EOS's
examined here, the presence of a kaon condensate postpones the transition to a
quark matter phase to higher densities.

The general conclusion which emerges from the above analysis is that the
softening provided by the occurence of kaon condensation invariably postpones
the transition to a phase comprising of quark matter to higher densities than
the case without kaon condensation.  The transition density and hence the size
of the quark core depends rather sensitively on whether the underlying nuclear
EOS is soft or stiff at the relevant densities, and on the bag
constant, $B$.

\bigskip
\bigskip
\addtocounter{section}{1}
\setcounter{equation}{0}
{\centerline {\bf 5. Neutrino Trapping }}
\bigskip
\bigskip

In a newly formed neutron star which is less than several seconds old,
neutrinos of all species are locally trapped in dense matter (density greater
than about $10^{13}$ g cm$^{-3}$).  In this case, the beta equilibrium
conditions in the matter are altered from what is described in sect. 2.
Thus, the composition of matter is affected, and the threshold densities for
both kaon condensation and  for quark--hadron phase transition will be altered.
Because of trapping, the numbers of leptons per baryon of each flavor of
neutrino,
\be
Y_{Le} = x_e + Y_{\nu_e}\,, \qquad
Y_{L\mu} = x_\mu + Y_{\nu_\mu}\,, \qquad
{\rm~and~} \qquad Y_{L\tau} = x_\tau + Y_{\nu_\tau}\,, \label{nleptons}
\ee
are conserved on beta equilibrium time scales.  As
neutrinos diffuse through the matter, the lepton numbers will change.  It is
beyond the scope of this work to make detailed predictions of how the lepton
numbers change with time, but it is interesting nevertheless to make some
observations regarding the effect of trapped neutrinos.

Neutron stars are thought to be formed as the result of the
gravitational collapse of the white--dwarf core of massive stars to
nuclear densities.  Beyond densities of about $10^{13}$ g cm$^{-3}$,
neutrinos are trapped in the matter, being unable to propagate on
dynamical timescales.  Electron capture reactions, which proceeded due
to the increasing density and electron chemical potential, effectively
halt as the trapped neutrinos settle into a degenerate Fermi sea. The
contribution to pressure from  neutrinos of a given species is
\be
P_{\nu_e} = (1/24\pi^2)\mu_{\nu_e}^4
\ee
The new beta equilibrium condition is
\be
\mu=\mu_n-\mu_p=\mu_e-\mu_{\nu_e},
\ee
where $\mu_{\nu_e}$, the chemical potential of the electron neutrinos,
is related to the net electron--neutrino number per baryon, $Y_{\nu_e}$ by
\be
\mu_{\nu_e}^3 = 6\pi^2\rho_0uY_{\nu_e}
\ee
For a given $Y_{\nu_e}$, the equilibrium electron fraction $x_e$ is obtained by
solving
\be
4(1-2x)S(u) = (3\pi^2\rho_0u)^{1/3}
\left[ x_e^{1/3}-\left(2(Y_{Le}-x_e)\right)^{1/3} \right] \,. \label{trape}
\ee
At the beginning of collapse, the electron lepton number
$Y_{Le}=x_e\simeq0.42$.  At the onset of trapping, it has declined to
$Y_{Le}=x_e+Y_{\nu_e}\simeq0.4$, the precise value depending on the efficiency
of electron capture reactions during the initial collapse stage.  As a
consequence, by the time the density $\rho_0$ is reached, one finds that
$x=x_e\simeq0.32$ and $\mu\simeq44$ MeV (with $F(u)=u$) in contrast to
the untrapped values of $x=x_e\simeq0.038$ and  $\mu\simeq111$ MeV.  The
decrease in $\mu$ is a result of the now--large value for $\mu_{\nu_e}$.
Depending on the form of the symmetry energy chosen, muons can appear at high
density.  The equilibrium lepton fractions are then determined by solving
eq.~(\ref{trape}) together with
\def\root{\left(1 + \frac {m_\mu^2}{(3\pi^2\rho x_\mu)^{2/3}}\right)^{1/2} }
\be
4(1-2x)S(u) = (3\pi^2\rho_0u)^{1/3}
\left[ x_\mu^{1/3}\root -\left(2(Y_{L_\mu}-x_\mu)\right)^{1/3} \right] \,.
\label{trapmu}
\ee
and the charge neutrality relation $x=x_e+x_\mu$. Initially, because
no muons are present when neutrinos become trapped, the additional
constraint $Y_{L\mu}=x_\mu+Y_{\nu_\mu}=0$ can been imposed.

In fig.~9, the lepton fractions $x_e$ and $Y_{\nu_e}$ are shown as a
function of the density ratio $u$, for the three choices of $F(u)$ in
eq.~(\ref{fu}).  The role of interactions may be gauged from the
results for non--interacting nucleons (dashed lines), for which $S(u)
= \left( 2^{\frac{2}{3}}-1 \right) \frac{3}{5}E_F^{(0)}
u^{\frac{2}{3}}$.  The more rapidly $S(u)$ increases with density, the lower
the density at which muons appear. For example, with $F(u)=u$, muons
appear at $u \sim 4$, while with the choice $F(u)={\sqrt u}$, muons
cannot appear till $u \sim 8$.  The presence of muons, however, has
very little effect on the electron lepton fractions (over the case
without the inclusion of muons), since $x_\mu$ remains extremely small
($\sim 10^{-4}$) over a wide range of densities.

The immediate consequence of smaller values for $\mu$ is to significantly {\it
delay} the onset of kaon (or pion) condensation until higher densities.  As in
the case for no trapped neutrinos, the threshold densities are dependent on the
functional forms of $S(u)$ but are independent of the overall stiffness of the
energy density function such as the form of $V(u)$ or the value of $K_0$.
Fig. 10 shows the dependence of the threshold density for kaon
condensation on the trapped electron neutrino number for two values of
$a_3m_s$.   The results shown correspond to the density at which
the relations in eq.~(\ref{thetamin}) (with $\theta = 0$),
eq.~(\ref{trape}), and eq.~(\ref{trapmu}) are simultaneously satisfied.
Clearly, the threshold density increases with $Y_{\nu_e}$.  This
feature of beta equilibrium with trapped neutrinos has important implications
for the evolution of young neutron stars.

When a neutron star is born, its central density is only about
$2\rho_0$.  This value is comparatively small because the
initial mass of the neutron star is no more than $1.0M_\odot$.  With
such small initial masses and central densities, our condensate model
predicts that as long as neutrinos are trapped, no condensate will
initially exist in the neutron star's core.

A neutron star acquires its final mass through accretion of matter
that fails to be ejected in the supernova explosion.  This accretion
can continue for the first several seconds of its life~\cite{Bethe93}.
Although the lepton concentrations in the newly formed star
are large and give rise to considerable lepton pressures, the loss of
lepton pressure is largely compensated by an increase in nuclear
symmetry pressure as the neutrinos diffuse out.  Figure 11 displays
the total mass of a neutron star as a function of density for the case
in which neither a kaon condensate nor a quark transition occurs.  Two
situations are displayed for each of two typical compression moduli,
$K_0$.  The first corresponds to trapped electron neutrinos
$(Y_{Le}=0.4)$, the initial state of a neutron star; the second to cold,
catalyzed matter $(Y_{\nu_e}=0.0)$, the final state of a neutron star.
Obviously, in the absence of mass accretion, the central density of a
star would, in general, not evolve much and could actually have a net
decrease with time. However, after the accretion of several
tenths of a solar mass of matter, the star's central density should
rise appreciably.

Thus, the central density of a young neutron star is expected to
increase as additional mass is accreted.  At the same time, the threshold
density for condensation will significantly decrease as neutrinos
diffuse out and $Y_{\nu_e}$ drops.  Thus, although not initially
present, a kaon condensate could appear during the evolution of
the star.  Figures 12 and 13 show the mass--central density relation
for neutron stars with kaon condensates for the cases $a_3m_s=-134$
MeV and $a_3m_s=-222$ MeV, respectively.  Two values for $K_0$, 180
and 240 MeV, and the cases of trapped neutrinos and cold, catalyzed
matter are shown.  These figures should be compared to figure 11 which
shows the identical cases when kaons do not condense.  The reduced
maximum mass in the case of a condensate is noteworthy, especially
when $K_0$ is small and $|a_3m_s|$ is large, as is the fact
that the evolution will invariably be to a much larger
central density.  This can be expected to significantly
influence the neutrino emission from newly formed neutron stars.

Similar arguments apply to the quark--hadron phase transition which will now
occur at higher densities than before.   The beta equilibrium condition
corresponding to eq.~(\ref{qbeta}) now reads
\be
\mu_d = \mu_u + \mu_e - \mu_{\nu_e} = \mu_s     \label{qtrap}
\ee

The equilibrium electron lepton fractions are obtained by  solving
simultaneously the eqs.~(\ref{qcharge}), (\ref{bnumber}) and  (\ref{qtrap})
together with the  constraint for the electron lepton number in
eq.~(\ref{nleptons}).  For non--interacting quarks, this amounts to
simultaneously solving the two equations:
\be
(1+x_e) + \left[ x_s^{2/3} + C/u^{2/3} \right]^{3/2} + x_s = 3 \nonumber \\
(1+x_e)^{1/3} - \left[ x_s^{2/3} + C/u^{2/3} \right]^{1/2} + (3x_e)^{1/3}
- \left[6(Y_{Le}-x_e)\right]^{1/3} = 0  \label{msbeta}
\ee
where the factor $C=m_s^2/\pi^{4/3}\rho_0^{2/3}$.  The role of the strange
quark mass is not very significant, and if it is neglected, one can
eliminate $x_s$ with the use of eq.~({\ref{msbeta}).  Then one obtains
\be
(1+x_e)^{1/3}  - (1-x_e/2)^{1/3} + (3x_e)^{1/3}
- \left[6(Y_{Le}-x_e)\right]^{1/3} = 0 \label{ynue}.
\ee
In the case when $Y_{Le}=0.4$, $x_e=0.233$ when all quarks are taken
to be massless, whereas with $m_s=200$ MeV, eq.~(\ref{msbeta}) yields
$x_e=0.249(0.246)$ for $u=5(8)$.  The sensitivity of $x_e$ to the
effects of exchange interactions is also small.  (To include exchange
interactions, divide the first two terms of eq.~(\ref{ynue}) by the
factor $(1-2\alpha_c/\pi)^{1/3}$).  For $\alpha_c=0.45(1)$, one
obtains $~x_e=0.229(0.222)$, indepedent of baryon density.  Note the
large magnitude of the electron fractions, which are significant
compared to the no--trapping result $x_e\simeq 0$.  The
pressure support provided by the more abundant electrons and the
degenerate anti--neutrinos has the overall effect of softening the
EOS, since, at a fixed baryon density, the total pressure is smaller
than the pressure without trapped neutrinos.

These scenarios are not just of academic interest.  With the advent of
neutrino detectors capable of observing thousands of neutrinos from a
typical supernova located within our galaxy~\cite{BKG}, it is becoming
feasible to test different scenarios observationally.  There are two
aspects of the evolution that will be very important from the point of
view of discrimination:  The total neutrino energy radiated and the
net excess of $\nu_e$ emission relative to other neutrino types.

The total neutrino energy radiated is determined by the final binding
energy of the star.  To be specific, figure 14 shows the binding
energy of 1.4 $M_\odot$ stars as a function of $K_0$ and $a_3m_s$.  In
the upper-left part of the figure, where the contours are horizontal,
there is no condensate because the central density of the star is too
small.  The central density of a 1.4 $M_\odot$ star without a
condensate is not very sensitive to the value of $K_0$, and, hence,
neither is the binding energy.  However, if a condensate is present,
the central density and binding energy rise abruptly.  In general,
stars of a given mass containing a condensate can have much larger
binding energies than those without.

Electron neutrinos escape from the core, where they have energies of
order 200-300 MeV, by diffusing to the surface where they are radiated
with energies of order 10--20 MeV.  In the process of diffusion they
lose the bulk of their energy which eventually appears as
$\nu\overline\nu$ pairs of all three neutrino flavors.  Thus each
electron neutrino generates about 3 $\nu\overline\nu$ pairs of each
flavor.  It is to be expected that although the bulk of the emitted
neutrinos from a cooling neutron star are equally distributed among
neutrino types of all species, a slight (10--20\%) excess of $\nu_e$'s
will exist due to the net change in $x_e$ and $Y_{\nu e}$ in the
star's interior.  The net excess of $\nu_e$'s will depend on the change
in the total electron-lepton number, $Y_{Le}=Y_{\nu_e}+x_e$, during
this stage.  The change in $Y_{Le}$ will be larger in the case of a
meson condensate because the final electron fraction will be much
smaller.  If there is no condensate, the final density will be about
$3\rho_o$ and the final electron fraction will be $x_e\simeq0.1$.  If
a condensate exists, the final density will be much larger and the
final electron fraction can even be negative.  Whether this will
translate into an observable increase in relative electron-neutrino
emission or not is the subject of ongoing studies.


Since the binding energy appears observationally as neutrino energy,
condensates will result in a larger net neutrino emission.  As
discussed previously, one does not expect the kaon condensate to exist
when the neutron star is born, but if it appears it should do so after
several seconds, the neutrino diffusion/mass accretion timescale.
This timing may be translated into a perceptible change in the
neutrino luminosity marking the condensate's appearance.  It is very
interesting that observations of neutrinos from neutron stars in formation 
may provide direct evidence for a condensate's formation.

In addition, due to the separate conservation of each lepton number on
dynamical timescales, variations in $x_\mu$ might also give rise to
observable consequences.  Initially, in both cases there are no muons
present due to the low value of $\mu$ caused by neutrino trapping.
However, as the $\nu_e$'s leak out of the core, $\mu$ increases and
muons eventually appear.  Conservation of muon lepton number then
implies that these muons must be accompanied by an equal number of
$\overline\nu_\mu$'s, which will begin to diffuse to the surface.
Thus, as the value of $x_\mu$ increases, one can anticipate an
excess in emitted $\overline\nu_\mu$'s relative to $\nu_\mu$'s.  In
the case of no condensate or quark core, this situation persists to
the end of the evolution, but in the case in which condensates and/or
a quark core appears, the value of $x_\mu$ drops and could even become
negative.  Once this occurs, there would have to be a reversal of the
previous trend, and a net excess of $\nu_\mu$'s would now be emitted.

Finally, consider the case in which a quark core forms.  The
quark--hadron transition density, which is shifted to higher values
both by trapped neutrinos and by the possible occurence of a
condensate, may never be achieved in a neutron star.  If it is
achieved, it will probably occur only after neutrinos have diffused
out of the star's core and the final mass of the star has accumulated,
a time of several seconds at least.  Quark cores have vanishingly small
electron concentrations.  Thus the net change in $Y_{Le}$ will not
be as large as in the case of a kaon condensate, although stars with
quark cores have larger binding than similarly massive stars without
quark cores.

Therefore, it may be possible to discriminate among these situations
solely on the basis of neutrino observations.

It remains to be seen if the predicted excesses of one neutrino type
over its antiparticle are potentially observable.  The excesses will
depend on details of neutrino transport in cooling neutron stars, a
subject which is beyond the scope of the present paper.  It should
also be pointed out that the composition and when the condensate/quark
core appears, if ever, will depend on the star's temperature, which
undergoes considerable evolution during this stage.  As electron
neutrinos diffuse out of the core, their degeneracy energy is left
behind, and the star will heat up~\cite{BL}.  Due to increases in both
density and entropy, the central temperature should eventualy exceed
50 MeV.  {}From previous work on the effects of finite temperatures on
pion condensation~\cite{KB}, it is known that such temperatures will
inhibit condensation formation. Whether a condensate is actually
prevented from appearing until the temperature eventually drops, after 10--20
seconds, or whether a
condensate can appear earlier, remains unclear and is the subject of
ongoing study.  The results of this section should apply to other
models of kaon and pion condensation with some generality.  Needless
to say, the prospect of observationally determining the existence or
absence of a meson condensate or quark core is an exciting one.

\bigskip\bigskip
\addtocounter{section}{1}
\setcounter{equation}{0}
{\centerline {\bf 6. Outlook }}
\bigskip
\bigskip

In closing, we turn to extensions and improvements that are promising
directions for future study.  We have not included hyperons, which may
enter at a few times nuclear matter density, that interact with kaons
and nucleons via p--wave interactions.  While expected to inhibit kaon
condensation by increasing the numbers of degree of freedom, the
nature of the hyperon--nucleon interaction may instead allow a condensate to
form more readily.  Inclusion of all members of the meson and
baryon octets, in the determination of both the critical density and the
equation of state, may be necessary.  Relativistic corrections to the nucleon
wavefunction can be important at the central densities under consideration and
should be included.  The extent to which the mean field results presented here
are altered by further quantum corrections (loop effects) is also unclear at
the present time.  The structure and composition of stars
containing Bose condensates and/or quark matter may be intermediate to the
idealizations employed here.

During the completion of this work, we have received a few
preprints~\cite{Mutoetal,BLRT} in which some advances along the lines mentioned
above have been reported.  These recent findings affirm the essential results
concerning the onset of kaon condensation presented in this paper.

We thank G. E. Brown, K. Kubodera and M. Rho for stimulating and beneficial
discussions.  This research was supported in part by the US Department of
Energy grants DE--FG02--88ER40388 and DE--FG02--87ER40317 at the State
University of New York at Stony Brook.

\newpage
\renewcommand{\theequation}{A-\arabic{equation}}
\setcounter{equation}{0}
\bigskip
\bigskip
{\centerline {\bf Appendix A }}
\bigskip
\bigskip
{\ni {KAON SELF--ENERGY AND THE CONDENSATION THRESHOLD}}
\bigskip
\bigskip

The critical density for kaon condensation may be determined in a relatively
simple manner by studying the self--energy of kaons propagating in
matter {\it before} the onset of condensation, following the approach of Baym
and Flowers ~\cite{baymflow}.  {}From the Lagrangian density eq.\ (\ref{knexp}),
the field equation for the kaon  condensate is
\be
( \mu^2 - m_K^2 + \nabla^2 )\langle K^- \rangle - J = 0
\ee
where $\mu$ is the kaon chemical potential.  The source of the kaon field, $J$,
is obtained from $\L^{int}$, the terms in eq.\ (\ref{knexp}) involving the
interactions of kaons with nucleons.  Explicitly,
\be
J &=& - \langle \frac{\delta {\cal L}^{int}}{ \delta K^+ } \rangle
- i \mu \langle \frac{\delta {\cal L}^{int}}{ \delta \dot{K}^+ } \rangle
\ee
{}From the Hamilton equations of motion and eq.\ (\ref{rhok}) it follows that the
variation of the ground state energy, $\epsilon$, due to a variation  $\delta
\rho_K$ in the condensate, is
\be
\delta \epsilon = \left( ( m_K^2 - \mu^2 ) \langle K^- \rangle + J \right)
\delta \langle K^+ \rangle + c.c.
+ \mu \delta \rho_K
\label{envar}
\ee
Just above threshold, we may expand the source $J$ to first order in $\langle
K^- \rangle$, and use
\be
\frac {\delta J}{\delta \langle K^- \rangle } = \Pi(\mu)
\label{selfen}
\ee
where $\Pi$ is the kaon self--energy in medium, evaluated at frequency $\omega
= \mu$. Near threshold, the change in energy density due the condensate is
thus
\be
\epsilon = - D^{-1}(\mu)v_K^2
\label{envar2}
\ee
where the (zero--momentum) kaon inverse propagator  is
\be
D^{-1}(\omega) &=& \omega^2 - m_K^2 - \Pi(\omega) \nonumber\\
&=& \omega^2 - m_K^2
+ u \rho_0\frac{1+x}{2f^2} \omega
- \frac{u \rho_0}{2f^2} \left( 2 a_1 x + 2 a_2 + 4 a_3 \right) m_s
\label{dm1}
\ee
Note that the inverse propagator (\ref{dm1}) is density dependent and that the
density dependence of $x$ is given by the conditions of $\beta$--equilibrium
and charge neutrality in the absence of a condensate. The threshold is the
density at which the inverse kaon propagator evaluated at $\mu$ has a pole;
i.e. $\mu$ is equal to the energy $\omega$ of a physical particle at threshold
\be
D^{-1}(\mu) = 0
\label{threshold}
\ee
The energy of zero--momentum physical particles as given by the poles of the
inverse propagator (\ref{dm1}) are
\be
\omega^{\mp}
=  &{\mp}& u \rho_0 \frac{1+x}{2f^2}  \nonumber\\
&+& \sqrt{ \frac14 \left( u \rho_0 \frac{1+x}{2f^2} \right)^2
+ m_K^2
+ \frac{u \rho_0 }{2f^2}(2a_1x+2a_2+4a_3)m_s }
\label{omega}
\ee
where $\omega^{\mp}$ is the energy of $K^{\mp}$, respectively.

In order to determine the threshold from eq.\ (\ref{threshold}), we note that
to lowest (zeroth) order in $v_K^2$, the chemical potential $\mu_0$ and proton
fraction $x_0$ prior to kaon condensation  follow from eqs.\  (\ref{betatheta})
and (\ref{cntheta}), and are given by
\be
4(1-2x_0)S(u) = \mu_0
\label{betaeq0}
\ee
and
\be
- x_0u \rho_0 + \frac{\mu_0^3}{3 \pi^2} + \eta( \mu_0 - m_\mu)
\frac{{p_\mu}_0^3}{3 \pi^2 } = 0
\label{cn0}
\ee
where ${p_\mu}_0 = \sqrt{ \mu_0^2 - m_\mu^2}$. Equation (\ref{betaeq0}) may be
solved for $\mu_0$ and the solution inserted into eq.\ (\ref{cn0}).  One must
then evaluate the roots of a fourth order polynomial in $x$.

Thus, to determine the threshold for condensation,  one simply determines the
density at which the charge chemical potential {\it in the absence of a
condensate} corresponds to the energy of a physical particle, i.e. $D^{-1}(
\mu_0 ) = 0$.  Contributions to the energy density from electrons, muons and
the symmetry energy, affect the threshold only in the sense that they determine
the chemical potential $\mu_0$.

\newpage


\newpage
\begin{center}
\bigskip
\bigskip
Table 1
\bigskip
\bigskip

\begin{tabular}{lrrr} \hline \hline
$F(u)$     &   $a_3m_s$(MeV)     &     $\,\,\,\,u_c\,\,\,$  \\ \hline
           &   - 134            &     4.18\\
$u$        &   - 222             &     3.08\\
           &   - 310             &     2.42\\ \hline
           &   - 134             &    3.80 \\
$\frac{2u^2}{1+u}$
           &   - 222             &    2.88 \\
           &   - 310             &    2.30\\ \hline
           &   -134            &    4.95\\
$\sqrt{u}$ &   -222            &    3.41\\
           &   -310             &    2.57 \\ \hline \hline
\end{tabular}

\begin{quote}
{The critical density for the various potential contributions to the symmetry
energy $F(u)$ in eq.(\ref{fu}) and values of the parameter $a_3m_s$ considered
in the text.}
\end{quote}
\label{critdens}
\end{center}
\newpage

\newpage
\begin{center}
\bigskip
\bigskip
Table 2
\bigskip
\bigskip

\begin{tabular}{ccccccc}
\hline \hline \\ [.1in]
$K_0$ & $A$ & $B$ & $B^{\prime}$ & $\sigma$ & $C_1$ & $C_2$
 \\[.1in] \hline \\ [.1in]
120 & 75.94 & $-30.88$ & 0 & $0.498$ & $-83.84$ & $23$ \\[.1in]
180 &  440.94 & $-213.41$ & 0 & $0.927$ & $-83.84$ & $23$ \\[.1in]
240 & $-46.65$ & 39.45 & 0.3 & 1.663 & $-83.84$ & $23$
\\[.1in]
\hline \hline

\end{tabular}
\begin{quote}
{Parameters in eq.~(\ref{vu}) determined by fitting the equilibrium properties
of symmetric nuclear matter for some input values of the compression
modulus $K_0$~\cite{pal}. All energies in MeV. The finite--range
parameters $\Lambda_1 = 1.5p_F^{(0)}$ and $\Lambda_2=3p_F^{(0)}$. }
\end{quote}
\label{palpar}
\end{center}
\newpage

\begin{center}
\bigskip
\bigskip
Table 3
\bigskip
\bigskip

\begin{tabular}{rrrrrrrr} \hline \hline
     u  &$\theta_{min}$&$\Delta \epsilon$&$\mu$&$x$&$x_K$&$x_e$&$x_\mu$ \\
\hline
   4.18 &   0.0 &    0.0 &   256.1 &  0.195 &  0.000 &  0.110 &  0.084\\
   4.68 &  28.9 &   -2.5 &   228.1 &  0.278 &  0.160 &  0.070 &  0.049\\
   5.18 &  40.6 &   -9.7 &   198.9 &  0.344 &  0.276 &  0.042 &  0.026\\
   5.68 &  48.9 &  -20.8 &   169.7 &  0.393 &  0.358 &  0.024 &  0.012\\
   6.18 &  55.0 &  -35.2 &   142.0 &  0.430 &  0.413 &  0.013 &  0.004\\
   6.68 &  59.5 &  -52.0 &   116.6 &  0.457 &  0.450 &  0.007 &  0.001\\
   7.18 &  62.8 &  -70.8 &    93.7 &  0.477 &  0.473 &  0.003 &  0.000\\
   7.68 &  65.3 &  -90.9 &    73.5 &  0.491 &  0.490 &  0.001 &  0.000\\
   8.18 &  67.3 & -112.2 &    55.0 &  0.502 &  0.501 &  0.001 &  0.000\\
   8.68 &  68.9 & -134.5 &    38.2 &  0.510 &  0.510 &  0.000 &  0.000\\
   9.18 &  70.1 & -157.4 &    24.0 &  0.516 &  0.516 &  0.000 &  0.000\\
   9.68 &  71.1 & -181.1 &    10.8 &  0.521 &  0.521 &  0.000 &  0.000\\
  10.18 &  71.9 & -205.2 &    -1.0 &  0.525 &  0.525 &  0.000 &  0.000\\
  10.68 &  72.5 & -229.8 &   -10.9 &  0.528 &  0.528 &  0.000 &  0.000\\
  11.18 &  73.1 & -254.9 &   -21.7 &  0.530 &  0.530 &  0.000 &  0.000\\
  11.68 &  73.5 & -280.3 &   -30.0 &  0.532 &  0.532 &  0.000 &  0.000\\
  12.18 &  73.8 & -306.1 &   -37.1 &  0.533 &  0.533 &  0.000 &  0.000\\
  12.68 &  74.1 & -332.2 &   -44.6 &  0.534 &  0.534 &  0.000 &  0.000\\
  13.18 &  74.4 & -358.7 &   -52.5 &  0.535 &  0.535 &  0.000  &  0.000\\
  \hline \hline
\end{tabular}

\begin{quote}
{Physical quantities as a function of density ratio $u$ for  $a_3m_s = -134$
MeV.  Tabulated are the angle $\theta_{min}$ (in degrees), energy density gain
$\Delta \epsilon = \epsilon(u,\theta) - \epsilon (u,0)$
(${\rm MeV}/{\rm fm}^3$), the chemical potential $\mu$ (MeV), the proton
fraction $x$, and the kaon fraction $x_K$, electron and muon fractions, $x_e$
and  $x_\mu$. }
\end{quote}
\label{theta1}
\end{center}
\newpage

\begin{center}
\bigskip
\bigskip
Table 4
\bigskip
\bigskip

\begin{tabular}{rrrrrrrr} \hline \hline
     u  &$\theta_{min}$&$\Delta \epsilon$&$\mu$&$x$&$x_K$&$x_e$&$x_\mu$ \\
\hline
   3.08 &   0.00 &  0.00  &   218.9 &  0.157 &  0.000 &  9.35(-2) &  6.28(-2) \\
   3.58 &  39.28 &  -5.6  &   162.0 &  0.323 &  0.277 &  3.26(-2) &  1.42(-2) \\
   4.08 &  56.08 &  -21.7 &    98.7 &  0.442 &  0.436 &  6.47(-3) &  0.000 \\
   4.58 &  67.34 &  -45.8 &    38.0 &  0.516 &  0.516 &  3.31(-4) &  0.000 \\
   5.08 &  74.76 &  -75.1 &   -12.8 &  0.558 &  0.558 & -1.13(-4) &  0.000 \\
   5.58 &  79.47 & -107.6 &   -53.5 &  0.580 &  0.580 & -7.54(-3) &  0.000 \\
   6.08 &  82.55 & -141.8 &   -86.4 &  0.591 &  0.594 & -2.92(-3) &  0.000 \\
   6.58 &  84.66 & -177.1 &  -113.7 &  0.597 &  0.604 & -6.13(-3) & -3.07(-4)\\
   7.08 &  86.26 & -213.2 &  -136.6 &  0.600 &  0.613 & -9.89(-3) & -2.52(-3) \\
   7.58 &  87.49 & -249.9 &  -156.3 &  0.602 &  0.621 & -1.38(-2) & -5.53(-3) \\
   8.08 &  88.44 & -286.9 &  -173.4 &  0.601 &  0.628 & -1.77(-2) & -8.83(-3) \\
   8.58 &  89.18 & -324.4 &  -188.5 &  0.600 &  0.634 & -2.14(-2) & -1.22(-2) \\
   9.08 &  89.76 & -362.1 &  -201.9 &  0.599 &  0.640 & -2.49(-2) & -1.54(-2) \\
   9.58 &  90.20 & -400.1 &  -214.0 &  0.598 &  0.645 & -2.81(-2) & -1.85(-2) \\
  10.08 &  90.54 & -438.3 &  -224.9 &  0.596 &  0.649 & -3.10(-2) & -2.13(-2) \\
  10.58 &  90.80 & -476.7 &  -234.9 &  0.595 &  0.652 & -3.37(-2) & -2.40(-2) \\
  11.08 &  90.99 & -515.2 &  -244.0 &  0.593 &  0.655 & -3.60(-2) & -2.64(-2) \\
  11.58 &  91.12 & -553.9 &  -252.4 &  0.591 &  0.658 & -3.81(-2) & -2.86(-2) \\
  12.08 &  91.21 & -592.8 &  -260.2 &  0.589 &  0.660 & -4.00(-2) & -3.06(-2)\\
  \hline \hline
\end{tabular}

\begin{quote}
{Physical quantities as a function of density ratio $u$ for  $a_3m_s = -222$
MeV.  Tabulated are the angle $\theta_{min}$ (in degrees), energy density gain
$\Delta \epsilon = \epsilon(u,\theta) - \epsilon (u,0)$
(${\rm MeV}/{\rm fm}^3$), the chemical potential $\mu$ (MeV), the proton
fraction $x$, and the kaon fraction $x_K$, electron and muon fractions, $x_e$
and  $x_\mu$. }  The numbers in parenthesis are exponents to the base 10.
\end{quote}
\label{theta2}
\end{center}
\newpage

\begin{center}
\bigskip
\bigskip
Table 5
\bigskip
\bigskip

\begin{tabular}{rrrrrrrr} \hline \hline
     u  &$\theta_{min}$&$\Delta \epsilon$&$\mu$&$x$&$x_K$&$x_e$&$x_\mu$ \\
\hline
   2.42 &   5.28 &   0.00 &   190.9 &  0.132 &  0.00  &  0.080 &  0.047\\
   2.92 &  53.69 &  -11.8 &    86.0 &  0.433 &  0.427 &  0.006 &  0.000\\
   3.42 &  80.50 &  -46.4 &   -32.0 &  0.606 &  0.607 &  0.000 &  0.000\\
   3.92 &  92.47 &  -93.9 &  -110.7 &  0.655 &  0.665 & -0.009 &  0.000\\
   4.42 &  98.76 & -146.2 &  -163.6 &  0.666 &  0.706 & -0.027 & -0.012\\
   4.92 & 102.81 & -200.4 &  -203.0 &  0.665 &  0.742 & -0.047 & -0.029\\
   5.42 & 105.68 & -255.8 &  -234.4 &  0.661 &  0.773 & -0.065 & -0.046\\
   5.92 & 107.81 & -311.8 &  -260.6 &  0.655 &  0.800 & -0.082 & -0.063 \\
   6.42 & 109.44 & -368.1 &  -282.9 &  0.649 &  0.823 & -0.097 & -0.077 \\
   6.92 & 110.72 & -424.7 &  -302.4 &  0.643 &  0.843 & -0.109 & -0.090 \\
   7.42 & 111.73 & -481.3 &  -319.8 &  0.637 &  0.860 & -0.121 & -0.102 \\
   7.92 & 112.53 & -538.0 &  -335.4 &  0.632 &  0.874 & -0.131 & -0.112 \\
   8.42 & 113.17 & -594.8 &  -349.5 &  0.626 &  0.886 & -0.139 & -0.121 \\
   8.92 & 113.67 & -651.4 &  -363.4 &  0.621 &  0.897 & -0.147 & -0.128 \\
   9.42 & 114.06 & -708.1 &  -374.3 &  0.618 &  0.905 & -0.153 & -0.135 \\
   9.92 & 114.36 & -764.6 &  -385.3 &  0.614 &  0.913 & -0.158 & -0.141 \\
  10.42 & 114.59 & -821.2 &  -395.5 &  0.610 &  0.919 & -0.163 & -0.146 \\
  10.92 & 114.74 & -877.6 &  -405.1 &  0.607 &  0.924 & -0.167 & -0.150 \\
  11.42 & 114.84 & -933.9 &  -414.1 &  0.603 &  0.928 & -0.171 & -0.154 \\
 \hline \hline
\end{tabular}

\begin{quote}
{Physical quantities as a function of density ratio $u$ for  $a_3m_s = -310$
MeV.  Tabulated are the angle $\theta_{min}$ (in degrees), energy density gain
$\Delta \epsilon = \epsilon(u,\theta) - \epsilon (u,0)$  (${\rm MeV}/{\rm
fm}^3$), the chemical potential $\mu$ (MeV), the proton fraction $x$, and the
kaon fraction $x_K$, electron and muon fractions, $x_e$ and  $x_\mu$. }
\end{quote}
\label{theta3}
\end{center}
\newpage

\begin{center}
\bigskip
\bigskip
Table 6
\bigskip
\bigskip

\begin{tabular}{cccccc} \hline \hline
$F(u)$ & $K_0$ & $\frac{M_{max}}{M_\odot}$
                & $R$
                        & $\frac{\rho_{cent}}{\rho_0}$
                        & $\Omega_K  $  \\
       & (MeV)  &         &  (km) & & $( 10^4 {\rm s}^{-1} )$ \\ \hline
       & 120    &  1.458  & 9.178 & 10.682 & 1.057 \\
$u$    & 180    &  1.723  & 9.922 & 8.855  & 1.023 \\
       & 240    &  1.937  & 10.565& 7.283  & 0.987 \\ \hline
       & 120    &  1.475  & 9.674 & 10.000 & 0.983 \\
$ \frac {2u^2}{ 1+ u} $
       & 180    &  1.739  & 10.327 & 8.149 & 0.968 \\
       & 240    &  1.953  & 10.938 & 6.947 & 0.941 \\ \hline
       & 120    &  1.405  &  8.423 & 12.306 & 1.181 \\
$\sqrt{u} $
       & 180    &  1.680  &  9.350 & 9.375 & 1.104 \\
       & 240    &  1.897  & 10.087 & 7.816 & 1.047 \\ \hline \hline
\end{tabular}

\begin{quote}
{ Neutron star properties in the normal state for the EOS from
ref.~\cite{pal}. }
\end{quote}
\label{normal}
\end{center}
\newpage

\begin{center}
\bigskip
\bigskip
Table 7
\bigskip
\bigskip

\begin{tabular}{cccccc} \hline \hline
$F(u)$ & $K_0$ & $\frac{M_{max}}{M_\odot}$
                & $R$
                        & $\frac{\rho_{cent}}{\rho_0}$
                        & $\Omega_K  $  \\
       & (MeV)  &         &  (km) & & $( 10^4 {\rm s}^{-1} )$ \\ \hline
       & 120    &  1.152  & 6.628 & 19.453 & 1.532 \\
$u$    & 180    &  1.428  & 7.975  & 14.063 & 1.292 \\
       & 240    &  1.752  & 11.622 & 5.882  & 0.813 \\ \hline
       & 120    &  0.877  & 10.181  & 9.479 & 0.702 \\
$ \frac {2u^2}{ 1+ u} $
       & 180    &  1.418  & 8.053  & 13.906 & 1.269 \\
       & 240    &  1.741  & 12.344 & 5.213  & 0.741 \\   \hline
       & 120    &  0.931  & 9.442 & 9.529 & 0.810 \\
$\sqrt{u} $
       & 180    &  1.448  &  8.220 & 12.813 & 1.243 \\
       & 240    &  1.782  & 10.820 & 6.719  & 0.913 \\ \hline \hline
\end{tabular}

\begin{quote}
{ Kaon condensed star properties for $a_3m_s = -134$ MeV. }
\end{quote}
\label{a134}
\end{center}
\newpage

\begin{center}
\bigskip
\bigskip
Table 8
\bigskip
\bigskip

\begin{tabular}{cccccc} \hline \hline
$a_3m_s$ & $K_0$ & $\frac{M_{max}}{M_\odot}$
                & $R$
                        & $\frac{\rho_{cent}}{\rho_0}$
                        & $\Omega_K  $  \\
(MeV)  & (MeV)  &         &  (km) & & $( 10^4 {\rm s}^{-1} )$ \\ \hline
-222   & 180    &  1.441  & 6.62  & 16.25 & 1.72 \\
       & 240    &  1.576  & 7.55  & 13.13 & 1.47  \\ \hline \hline
\end{tabular}

\begin{quote}
{ Kaon condensed star properties for $a_3m_s = -222 $ MeV. }
\end{quote}
\label{soft}
\end{center}
\newpage

\section*{Figure Captions}

\hspace{0.25 in} Fig. 1.  Critical density determined from kaon self--energy.
The abscissa shows the density ratio $u=\rho/\rho_0$, with $\rho_0=0.16~{\rm
fm}^{-3}$.
Lower panel: $\omega^-(u)$ (solid line), $\mu_0(u)$ (long dashes) and
$\omega^+(u)$ (short dashes).  The critical density $u_c$ is given by the
intersection of $\omega^-(u)$ with $\mu_0(u)$. The
potential contribution to the symmetry energy is $F(u)=u$ and $a_3m_s = -134$
MeV.  Upper panel : same as lower panel but for $a_3m_s = -310$ MeV.

Fig. 2.  Two equations of state from ref.~\cite{pal} with compression modulus
$K_0 = 240 $ MeV and potential contribution to the symmetry energy $F(u) =
u=\rho/\rho_0$, with $\rho_0=0.16~{\rm fm}^{-3}$. Lower panel: EOS in absence
(short dashes) and presence (solid line) of a kaon condensate with $a_3m_s =
-134$ MeV.  Upper panel: EOS in absence (short dashes) and presence of a kaon
condensate before (long dashes) and after  (solid line) applying the Maxwell
construction; $a_3m_s = -222$ MeV.

Fig. 3.  Mass curves for $a_3m_s = -134$ MeV and $F(u)=u$.  Lower panel: Mass
vs.\  central density.  {}From the uppermost curve to the lowest, the compression
modulus is $K_0 = 240 $ MeV, $K_0 = 180 $ MeV and $K_0 = 120 $ MeV.  Upper
panel:  Mass vs. radius for the same values of $K_0$.

Fig. 4.  Mass curves for $a_3m_s = -222$ MeV and $F(u)=u$. Lower panel: Mass
vs.\ central density.  For the upper curve, the compression modulus is $K_0 =
240 $ MeV, for the lower, $K_0 = 180 $ MeV. Upper panel:  Mass vs. radius for
the same values of $K_0$.

Fig. 5. Contours of maximum mass of neutron stars as functions of
nuclear incomressibility ($K_0$) and the strength of the kaon
condensate ($a_3m_s$).  In the hatched region, no Maxwell construction
for the condensate was possible because $u_{min}<1$.  The region in
which the maximum mass was less than the binary pulsar constraint of
\sol is also shown.

Fig. 6.  Chemical potential versus pressure in the quark (solid curves) and
hadronic phases (dashed curves).  The quark phase EOS  was calculated following
ref.~\cite{FM} (see text for differences in input values). The lower (upper)
dashed curve in each panel refers to matter with (without) a kaon condensate.
The baryonic EOS~\cite{pal} in the  lower panel is stiffer (PAL31: $K_0=240$
MeV and $F(u)=u$) than that in the  upper panel(PAL21: $K_0=180$ MeV and
$F(u)=u$). The points show baryon density $u=\rho/ 0.16~{\rm fm}^{-3}$ in steps
of unity starting from $u=2$ in both phases.  The intersection of the solid and
dashed curves gives the onset of transition to quark matter.                   

Fig. 7.  Same as fig. 6 but for the quark EOS from ref.~\cite{FJ} with
$\alpha_c=0.45$ independent of baryon density.

Fig. 8. Same as fig. 6 but for the quark EOS with $\alpha_c=0.55/u^{2/3}$ for
the upper solid curve and  $\alpha_c=0.275/u^{1/3}$ for the lower solid curve.
The density dependence of $\alpha_c$ is after ref.~\cite{BBC}.

Fig. 9. Electron lepton fractions in the trapped neutrino regime.  The
solid curves are for interacting nucleons with $F(u)=u$ (open squares),
$F(u)=2u^2/1+u$ (filled squares) and $F(u)={\sqrt u}$ (crosses).  The
dashed curves are for non--interacting nucleons.

Fig. 10. Critical density for kaon condensation versus trapped neutrino number
per baryon.

Fig. 11. Mass versus central density with (solid curves) and without (dashed
curves) trapped neutrinos for normal stars.  For the upper (lower)
panel: $K_0 = 240(180)$ MeV. In both cases, $F(u) = u$.

Fig. 12. Mass versus central density with (solid curves) and without (dashed
curves) trapped neutrinos for kaon condensed stars ($a_3m_s=-134$ MeV).  For
the upper (lower) panel: $K_0 = 240(180)$ MeV. In both cases, $F(u) = u$.

Fig. 13. Mass versus central density with (solid curves) and without (dashed
curves) trapped neutrinos for kaon condensed stars ($a_3m_s=-222$ MeV).  For
the upper (lower) panel: $K_0 = 240(180)$ MeV. In both cases, $F(u) = u$.

Fig. 14. Contours of neutron star binding energy as a function of
nuclear incompressibility ($K_0$) and condensate strength ($a_3m_s$).  In
the hatched region, no Maxwell construction for the condensate was
possible because $u_{min}<1$.  The region in which the maximum mass
was less than the binary pulsar constraint of \sol is also shown.

\end{document}